\pdfoutput=1
\documentclass[12pt,a4paper]{article}

\usepackage{ifthen} 
\newboolean{pdflatex}
\setboolean{pdflatex}{true} 

\newboolean{articletitles}
\setboolean{articletitles}{true} 

\newboolean{uprightparticles}
\setboolean{uprightparticles}{false} 

\newboolean{inbibliography}
\setboolean{inbibliography}{false} 

\def\paperauthors{LHCb collaboration} 
\def\paperasciititle{Evidence for a new structure \\ in the J/psi p and J/psi pbar systems \\ in BsJ/psi p pbar decays} 
\def\papertitle{Evidence for a new structure \\ in the $J/\psi p$ and $J/\psi \bar{p}$ systems \\ in $B_s^0 \to J/\psi p \bar{p}$ decays} 
\def\paperkeywords{{High Energy Physics}, {LHCb}} 
\def\papercopyright{\the\year\ CERN for the benefit of the LHCb collaboration} 
\def\paperlicence{CC BY 4.0 licence}
\def\paperlicenceurl{https://creativecommons.org/licenses/by/4.0/}

\usepackage[top=1in, bottom=1.25in, left=1in, right=1in]{geometry}

\usepackage{microtype}
\usepackage{lineno}  
\usepackage{xspace} 
\usepackage{caption} 

\usepackage{multirow}
\usepackage{graphicx}  
\usepackage{color}
\usepackage{colortbl}
\graphicspath{{./figs/}} 

\usepackage{amsmath} 
\usepackage{amssymb}
\usepackage{amsfonts}
\usepackage{upgreek} 

\newcommand*\patchAmsMathEnvironmentForLineno[1]{%
\expandafter\let\csname old#1\expandafter\endcsname\csname #1\endcsname
\expandafter\let\csname oldend#1\expandafter\endcsname\csname
end#1\endcsname
 \renewenvironment{#1}%
   {\linenomath\csname old#1\endcsname}%
   {\csname oldend#1\endcsname\endlinenomath}%
}
\newcommand*\patchBothAmsMathEnvironmentsForLineno[1]{%
  \patchAmsMathEnvironmentForLineno{#1}%
  \patchAmsMathEnvironmentForLineno{#1*}%
}
\AtBeginDocument{%
\patchBothAmsMathEnvironmentsForLineno{equation}%
\patchBothAmsMathEnvironmentsForLineno{align}%
\patchBothAmsMathEnvironmentsForLineno{flalign}%
\patchBothAmsMathEnvironmentsForLineno{alignat}%
\patchBothAmsMathEnvironmentsForLineno{gather}%
\patchBothAmsMathEnvironmentsForLineno{multline}%
\patchBothAmsMathEnvironmentsForLineno{eqnarray}%
}

\usepackage{hyperxmp}

\usepackage[pdftex,
            pdfauthor={\paperauthors},
            pdftitle={\paperasciititle},
            pdfkeywords={\paperkeywords},
            pdfcopyright={Copyright (C) \papercopyright},
            pdflicenseurl={\paperlicenceurl}]{hyperref}

\usepackage[colorinlistoftodos,textsize=scriptsize]{todonotes}

\usepackage[bottom,flushmargin,hang,multiple]{footmisc}

\usepackage[all]{hypcap} 

\usepackage{xspace} 
\usepackage{upgreek}

\def\lhcb   {\mbox{LHCb}\xspace}

\def\MagUp {\mbox{\em Mag\kern -0.05em Up}\xspace}

\ifthenelse{\boolean{uprightparticles}}%
{

 \def\Pmu         {\ensuremath{\upmu}\xspace}

 \def\Ppi         {\ensuremath{\uppi}\xspace}

 \def\Ppsi        {\ensuremath{\uppsi}\xspace}

 \def\PDelta      {\ensuremath{\Delta}\xspace}                 
 \def\PXi         {\ensuremath{\Xi}\xspace}                 
 \def\PLambda     {\ensuremath{\Lambda}\xspace}                 
 \def\PSigma      {\ensuremath{\Sigma}\xspace}                 
 \def\POmega      {\ensuremath{\Omega}\xspace}                 
 \def\PUpsilon    {\ensuremath{\Upsilon}\xspace}

 \def\PB      {\ensuremath{\mathrm{B}}\xspace}                 
                  
 \def\PD      {\ensuremath{\mathrm{D}}\xspace}

 \def\PJ      {\ensuremath{\mathrm{J}}\xspace}                 
 \def\PK      {\ensuremath{\mathrm{K}}\xspace}

 \def\Pb      {\ensuremath{\mathrm{b}}\xspace}                 
 \def\Pc      {\ensuremath{\mathrm{c}}\xspace}

 \def\Pi      {\ensuremath{\mathrm{i}}\xspace}

 \def\Pp      {\ensuremath{\mathrm{p}}\xspace}

 \def\Ps      {\ensuremath{\mathrm{s}}\xspace}

 \def\thebaroffset{0.0em}
}
{

 \def\Pmu         {\ensuremath{\mu}\xspace}

 \def\Ppi         {\ensuremath{\pi}\xspace}

 \def\Ppsi        {\ensuremath{\psi}\xspace}                 
                  
 \mathchardef\PDelta="7101
 \mathchardef\PXi="7104
 \mathchardef\PLambda="7103
 \mathchardef\PSigma="7106
 \mathchardef\POmega="710A
 \mathchardef\PUpsilon="7107
                  
 \def\PB      {\ensuremath{B}\xspace}                 
                  
 \def\PD      {\ensuremath{D}\xspace}

 \def\PJ      {\ensuremath{J}\xspace}                 
 \def\PK      {\ensuremath{K}\xspace}

 \def\Pb      {\ensuremath{b}\xspace}                 
 \def\Pc      {\ensuremath{c}\xspace}

 \def\Pi      {\ensuremath{i}\xspace}

 \def\Pp      {\ensuremath{p}\xspace}

 \def\Ps      {\ensuremath{s}\xspace}

 \def\thebaroffset{0.18em}
}
\newcommand{\offsetoverline}[2][\thebaroffset]{\kern #1\overline{\kern -#1 #2}}%

\makeatletter
\ifcase \@ptsize \relax
  \newcommand{\miniscule}{\@setfontsize\miniscule{4}{5}}
\or
  \newcommand{\miniscule}{\@setfontsize\miniscule{5}{6}}
\or
  \newcommand{\miniscule}{\@setfontsize\miniscule{5}{6}}
\fi
\makeatother

\DeclareRobustCommand{\optbar}[1]{\shortstack{{\miniscule (\rule[.5ex]{1.25em}{.18mm})}
  \\ [-.7ex] $#1$}}

\def\muon       {{\ensuremath{\Pmu}}\xspace}
\def\mup        {{\ensuremath{\Pmu^+}}\xspace}
\def\mun        {{\ensuremath{\Pmu^-}}\xspace} 

\def\mumu       {{\ensuremath{\Pmu^+\Pmu^-}}\xspace}

\def\squark    {{\ensuremath{\Ps}}\xspace}

\def\cquark    {{\ensuremath{\Pc}}\xspace}

\def\bquark    {{\ensuremath{\Pb}}\xspace}

\def\pion   {{\ensuremath{\Ppi}}\xspace}

\def\pip    {{\ensuremath{\pion^+}}\xspace}

\def\kaon    {{\ensuremath{\PK}}\xspace}

\def\KorKbar {\kern \thebaroffset\optbar{\kern -\thebaroffset \PK}{}\xspace}

\def\Kp      {{\ensuremath{\kaon^+}}\xspace}
\def\Km      {{\ensuremath{\kaon^-}}\xspace}

\def\Dbar    {{\ensuremath{\offsetoverline{\PD}}}\xspace}
\def\D       {{\ensuremath{\PD}}\xspace}

\def\DorDbar {\kern \thebaroffset\optbar{\kern -\thebaroffset \PD}\xspace}

\def\Dp      {{\ensuremath{\D^+}}\xspace}
\def\Dm      {{\ensuremath{\D^-}}\xspace}

\def\DpDm    {\ensuremath{\Dp {\kern -0.16em \Dm}}\xspace}

\def\Dstarb  {{\ensuremath{\Dbar{}^*}}\xspace}

\def\B       {{\ensuremath{\PB}}\xspace}
\def\Bbar    {{\ensuremath{\offsetoverline{\PB}}}\xspace}

\def\BorBbar {\kern \thebaroffset\optbar{\kern -\thebaroffset \PB}\xspace}

\def\Bd      {{\ensuremath{\B^0}}\xspace}

\def\BdorBdbar {\kern \thebaroffset\optbar{\kern -\thebaroffset \Bd}\xspace}

\def\Bs      {{\ensuremath{\B^0_\squark}}\xspace}
\def\Bsbar      {{\ensuremath{{\overline {\B}^0_\squark }}}\xspace}
\def\Bsb     {{\ensuremath{\Bbar{}^0_\squark}}\xspace}
\def\BsorBsbar {\kern \thebaroffset\optbar{\kern -\thebaroffset \Bs}\xspace}

\def\Bds     {{\ensuremath{\B_{(\squark)}^0}}\xspace}

\def\jpsi     {{\ensuremath{{\PJ\mskip -3mu/\mskip -2mu\Ppsi}}}\xspace}

\def\Y#1S{\ensuremath{\PUpsilon{(#1S)}}\xspace}

\def\proton      {{\ensuremath{\Pp}}\xspace}
\def\antiproton  {{\ensuremath{\overline \proton}}\xspace}

\def\Lz          {{\ensuremath{\PLambda}}\xspace}

\def\LorLbar     {\kern \thebaroffset\optbar{\kern -\thebaroffset \PLambda}\xspace}
\def\Lambdares   {{\ensuremath{\PLambda}}\xspace}

\def\Sigmares    {{\ensuremath{\PSigma}}\xspace}

\def\Xires       {{\ensuremath{\PXi}}\xspace}

\def\Lc          {{\ensuremath{\Lz^+_\cquark}}\xspace}

\def\Xic         {{\ensuremath{\Xires_\cquark}}\xspace}

\def\Lb           {{\ensuremath{\Lz^0_\bquark}}\xspace}

\def\Xibm         {{\ensuremath{\Xires^-_\bquark}}\xspace}

\newcommand{\decay}[2]{\ensuremath{#1\!\to #2}\xspace} 

\def\to                 {\ensuremath{\rightarrow}\xspace}

\def\CP                {{\ensuremath{C\!P}}\xspace}

\def\BsJpp    {\decay{\Bs}{\jpsi\proton\antiproton}}
\def\BJpp    {\decay{\Bd}{\jpsi\proton\antiproton}}

\def\AT#1     {\ensuremath{A_{\mathrm{T}}^{#1}}\xspace}           

\def\C#1      {\ensuremath{\mathcal{C}_{#1}}\xspace}                       
\def\Cp#1     {\ensuremath{\mathcal{C}_{#1}^{'}}\xspace}                    
\def\Ceff#1   {\ensuremath{\mathcal{C}_{#1}^{\mathrm{(eff)}}}\xspace}        
\def\Cpeff#1  {\ensuremath{\mathcal{C}_{#1}^{'\mathrm{(eff)}}}\xspace}       
\def\Ope#1    {\ensuremath{\mathcal{O}_{#1}}\xspace}                       
\def\Opep#1   {\ensuremath{\mathcal{O}_{#1}^{'}}\xspace}                    

\newcommand{\ket}[1]{\ensuremath{|#1\rangle}}              
\newcommand{\braket}[2]{\ensuremath{\langle #1|#2\rangle}} 

\newcommand{\aunit}[1]{\ensuremath{\text{\,#1}}}       

\newcommand{\tev}{\aunit{Te\kern -0.1em V}\xspace}
\newcommand{\gev}{\aunit{Ge\kern -0.1em V}\xspace}
\newcommand{\mev}{\aunit{Me\kern -0.1em V}\xspace}
\newcommand{\kev}{\aunit{ke\kern -0.1em V}\xspace}
\newcommand{\ev}{\aunit{e\kern -0.1em V}\xspace}
 
\newcommand{\mevc}{\ensuremath{\aunit{Me\kern -0.1em V\!/}c}\xspace}
\newcommand{\gevc}{\ensuremath{\aunit{Ge\kern -0.1em V\!/}c}\xspace}
\newcommand{\mevcc}{\ensuremath{\aunit{Me\kern -0.1em V\!/}c^2}\xspace}
\newcommand{\gevcc}{\ensuremath{\aunit{Ge\kern -0.1em V\!/}c^2}\xspace}

\def\fb   {\ensuremath{\aunit{fb}}\xspace}
\def\invfb   {\ensuremath{\fb^{-1}}\xspace}

\newcommand{\chisq}{\ensuremath{\chi^2}\xspace}

\def\deriv {\ensuremath{\mathrm{d}}}

\def\gsim{{~\raise.15em\hbox{$>$}\kern-.85em
          \lower.35em\hbox{$\sim$}~}\xspace}
\def\lsim{{~\raise.15em\hbox{$<$}\kern-.85em
          \lower.35em\hbox{$\sim$}~}\xspace}

\def\PDF {PDF\xspace}

\def\pt         {\ensuremath{p_{\mathrm{T}}}\xspace}

\def\ptot       {\ensuremath{p}\xspace}

\def\evtgen     {\mbox{\textsc{EvtGen}}\xspace}

\def\geant      {\mbox{\textsc{Geant4}}\xspace}

\def\photos     {\mbox{\textsc{Photos}}\xspace}

\def\pythia     {\mbox{\textsc{Pythia}}\xspace}

\def\tell1  {TELL1\xspace}
\def\ukl1   {UKL1\xspace}

\newcommand{\eg}{\mbox{\itshape e.g.}\xspace}
\newcommand{\ie}{\mbox{\itshape i.e.}\xspace}

\usepackage{cite} 
\usepackage{mciteplus}

\usepackage{longtable} 
\usepackage{comment}

\begin{document}

\renewcommand{\thefootnote}{\fnsymbol{footnote}}
\setcounter{footnote}{1}



\begin{titlepage}
\pagenumbering{roman}

\vspace*{-1.5cm}
\centerline{\large EUROPEAN ORGANIZATION FOR NUCLEAR RESEARCH (CERN)}
\vspace*{1.5cm}
\noindent
\begin{tabular*}{\linewidth}{lc@{\extracolsep{\fill}}r@{\extracolsep{0pt}}}
\ifthenelse{\boolean{pdflatex}}
{\vspace*{-1.5cm}\mbox{\!\!\!\includegraphics[width=.14\textwidth]{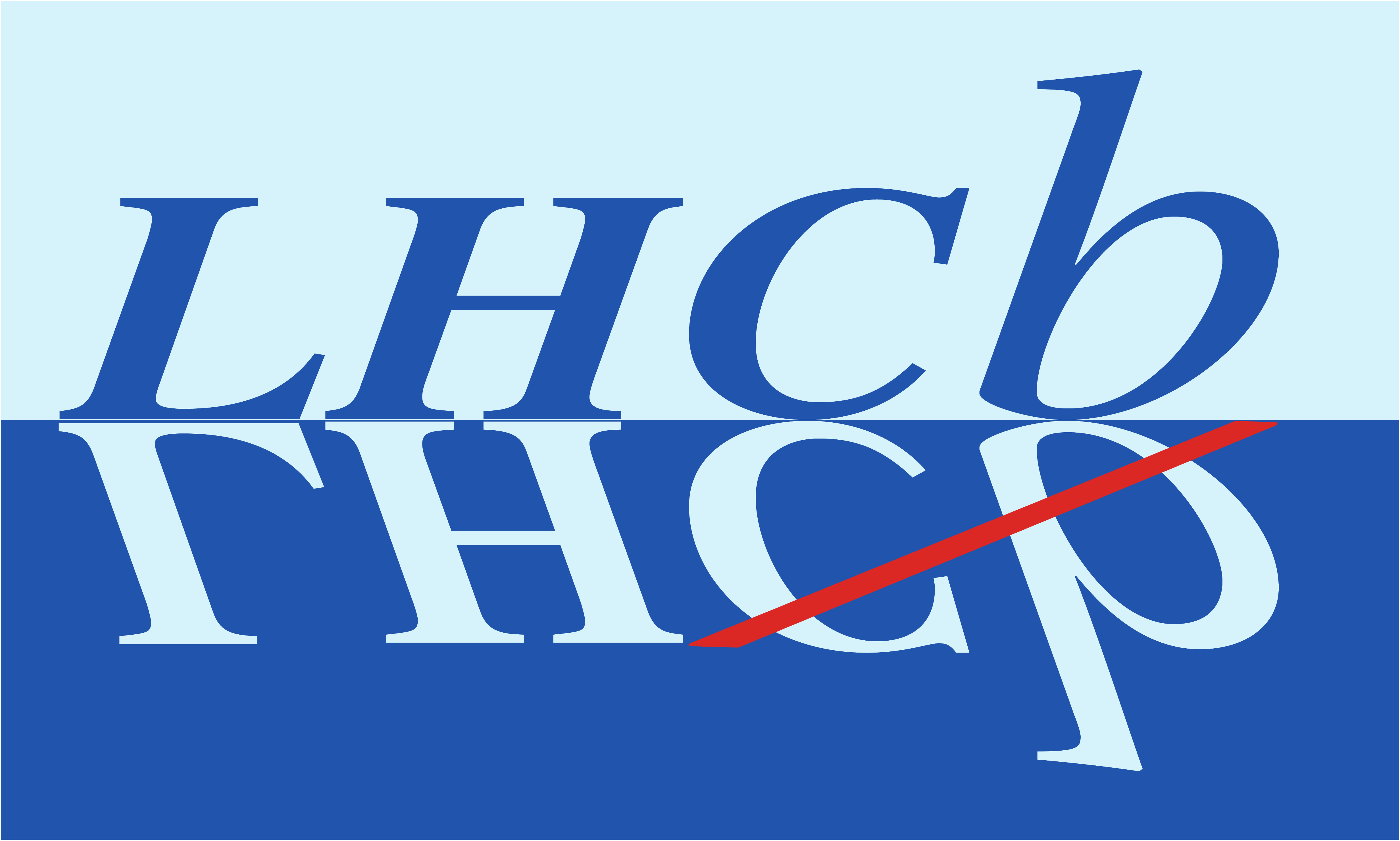}} & &}%
{\vspace*{-1.2cm}\mbox{\!\!\!\includegraphics[width=.12\textwidth]{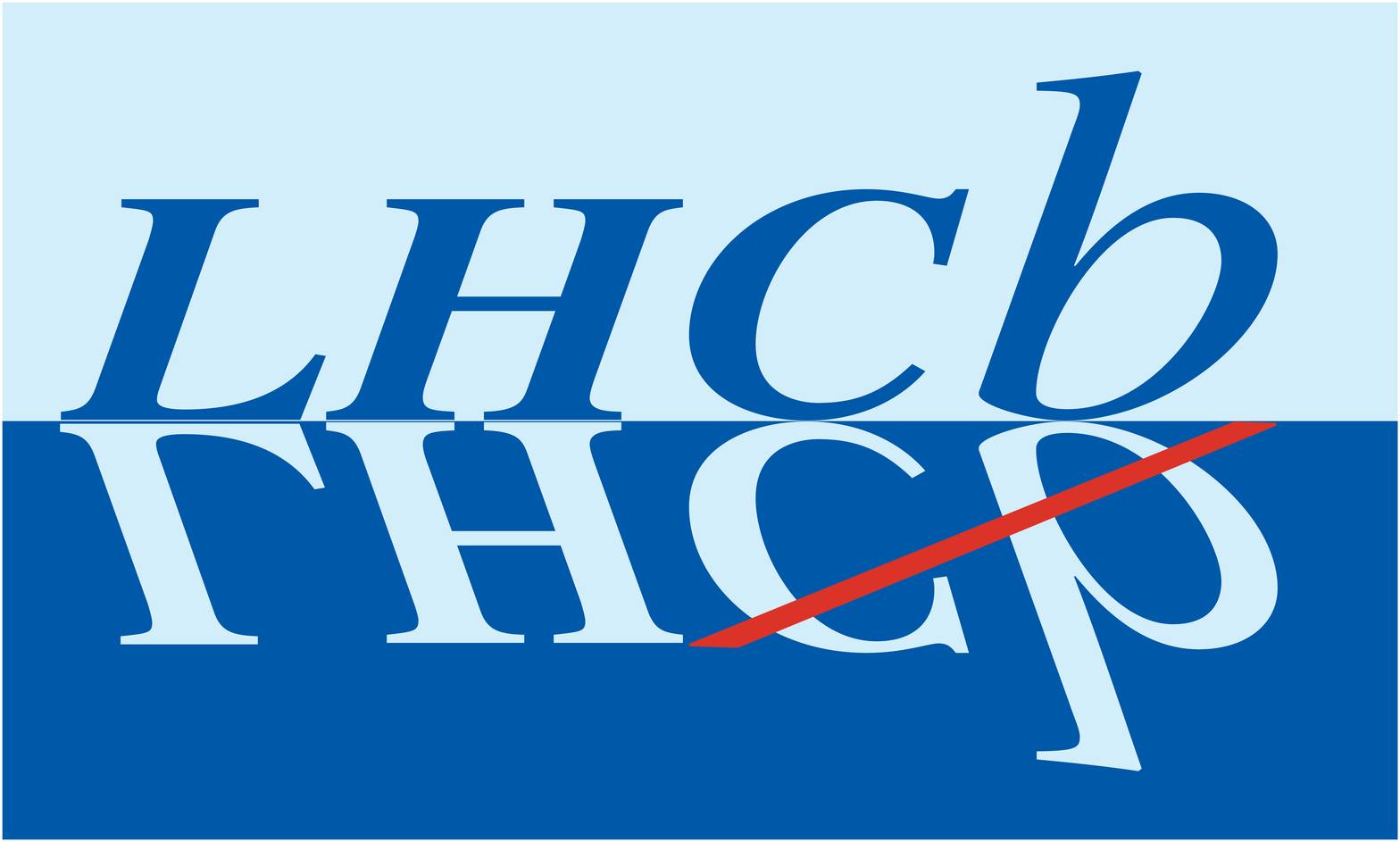}} & &}%
\\
 & & CERN-EP-2021-150 \\  
 & & LHCb-PAPER-2021-018 \\  
 & & February 7, 2022 \\ 
\end{tabular*}

\vspace*{4.0cm}

{\normalfont\bfseries\boldmath\huge
\begin{center}
  \papertitle 
\end{center}
}

\vspace*{2.0cm}

\begin{center}
\paperauthors\footnote{Authors are listed at the end of this Letter.}
\end{center}

\vspace{\fill}

\begin{abstract}
  \noindent
   An amplitude analysis of flavour-untagged $B_s^0 \to J/\psi p \bar{p}$ decays is performed using a sample of $797\pm31$ decays reconstructed with the LHCb detector. The data, collected in proton-proton collisions between 2011 and 2018, correspond to an integrated luminosity of 9 $\text{fb}^{-1}$. Evidence for a new structure in the $J/\psi p$ and $J/\psi \bar{p}$ systems with a mass of $4337 \ ^{+7}_{-4} \ ^{+2}_{-2}~\text{MeV}$ and a width of $29 \ ^{+26}_{-12} \ ^{+14}_{-14}~\text{MeV}$ is found, where the first uncertainty is statistical and the second systematic,  with a significance in the range of 3.1 to 3.7 $\sigma$, depending on the assigned $J^P$ hypothesis.
  
\end{abstract}

\vspace*{2.0cm}

\begin{center}
  Published in PhysRevLett.128.062001 

\end{center}

\vspace{\fill}

{\footnotesize 
\centerline{\copyright~\papercopyright. \href{\paperlicenceurl}{\paperlicence}.}}
\vspace*{2mm}

\end{titlepage}


\newpage
\setcounter{page}{2}
\mbox{~}


\renewcommand{\thefootnote}{\arabic{footnote}}
\setcounter{footnote}{0}

\cleardoublepage


\pagestyle{plain} 
\setcounter{page}{1}
\pagenumbering{arabic}


The observation of pentaquark candidates ($P_c$) in $\jpsi \proton$ final states produced in $\Lb \rightarrow \jpsi \proton \Km$ decays~\footnote{The charge-conjugate decay is implied, unless otherwise stated, and natural units  with $\hbar = c = 1$ are used throughout the paper.}~\cite{LHCb-PAPER-2015-029,LHCb-PAPER-2019-014} by the \lhcb experiment has stimulated interest in exotic spectroscopy. Recently, evidence for a  structure in the $\jpsi \Lambdares$ invariant-mass spectrum, consistent with a charmonium-like pentaquark with strangeness, was found in  ${\Xibm\rightarrow \jpsi\Lambdares \Km}$ decays~\cite{LHCb-PAPER-2020-039}. The mass of these states is just below threshold for the joint production of a charm baryon and a charm meson, \ie the $\Sigmares_c \Dstarb$ and the $\Xic\Dstarb$ thresholds for the $\jpsi \proton$ and the $\jpsi \Lz$ resonances, respectively. The mass separation from these thresholds might provide useful information for the phenomenological interpretation for these states. Proposed interpretation can be grouped into three classes: QCD-inspired models~\cite{Esposito:2016noz,Richard:2016eis}, 
residual hadron-hadron interaction models~\cite{Guo:2017jvc}
and rescattering effects particle~\cite{Guo:2019twa}. Additional measurements in different productions and decay channels 
are crucial to disentangle the various models~\cite{RevModPhys.90.015003}.

The \BsJpp decay was observed for the first time by the \lhcb experiment in 2019~\cite{LHCb-PAPER-2018-046}. 
This channel may have sensitivity to the resonant $P_c$ structures~\cite{LHCb-PAPER-2015-029,LHCb-PAPER-2019-014} within the $\jpsi \proton$ invariant-mass range of $[4034,4429]\mev$. 
Additionally, it could proceed via an intermediate glueball candidate $f_J(2220)$ decaying to $\proton \antiproton$~\cite{Hsiao:2014tda}.
Unlike $\Lb \rightarrow \jpsi \proton \Km$ decays receiving a relatively large contribution from the intermediate excited $\Lz$ resonances, no conventional states are expected to be produced in the \Bs decay, offering a clean environment to search for new resonant structures. 
Baryonic \Bds decays also allow for a study of the dynamics of the baryon-antibaryon system and its characteristic threshold enhancement, the origin of which is still
to be understood~\cite{Rosner_2003}.

In this Letter, an amplitude analysis of \BsJpp decay is presented, including a search for pentaquark and glueball states, using proton-proton ($pp$) collision data at centre-of-mass energies of 7\tev, 8\tev and 13\tev, corresponding to a luminosity of 9\invfb, collected between 2011 and 2018. The measurement is performed untagged, such that decays of \Bs and \Bsb are not distinguished and analysed together. 

The \lhcb detector is a single-arm forward
spectrometer covering the \mbox{pseudorapidity} range $2<\eta <5$, described in detail in Refs.~\cite{LHCb-DP-2014-002, LHCb-DP-2014-001, LHCb-DP-2013-003, LHCb-DP-2012-002}.
The online event selection is performed by a trigger~\cite{LHCb-DP-2012-004}, comprising a hardware stage based on information from the muon system which selects $\jpsi\to \mup\mun$ decays, followed by a software stage that applies a full event reconstruction. The software trigger relies on identifying \jpsi decays into muon pairs consistent with originating from a \B meson decay vertex detached from the primary $pp$ collision point.

Samples of simulated events are used to study the properties of the signal and control channels.  
The $pp$ collisions are generated using
\pythia~\cite{Sjostrand:2007gs} with a specific \lhcb
configuration~\cite{LHCb-PROC-2010-056}. Decays of hadronic particles and interactions with the detector material are described by \evtgen~\cite{Lange:2001uf}, using \photos~\cite{davidson2015photos}, and by the \geant toolkit~\cite{Allison:2006ve, *Agostinelli:2002hh, LHCb-PROC-2011-006}, respectively. The signal \BsJpp decays are generated from a uniform phase space distribution, while the $\Bs \to \jpsi\phi(\to \Kp\Km)$ control mode is generated according to the model of Ref.~\cite{LHCb-PAPER-2014-059}.

The event selection follows the same strategy as Ref. \cite{LHCb-PAPER-2018-046}.  
Signal \Bs candidates are formed from two pairs of oppositely charged tracks. The first pair is required to be consistent with muons originating from a \jpsi meson with a decay vertex significantly displaced from its associated primary $pp$ vertex (PV). For a given particle, the associated PV is the one with the smallest impact parameter $\chi^2_{\rm IP}$, defined as the difference in the vertex-fit \chisq of a given PV
 reconstructed with and without the track under consideration. 
The second pair is required to be consistent with protons originating from the muon-pair vertex. A kinematic fit~\cite{Hulsbergen:2005pu} to the $\Bs$ candidate is performed, with the dimuon mass constrained to the known \jpsi mass~\cite{PDG20}. 
The selection is optimised using multivariate techniques~\cite{Breiman} trained with simulation and data. Simulated events are weighted such that the distributions of momentum, $\ptot$, transverse momentum, $\pt$, and number of tracks per event for \Bs candidates match the $\Bs \to \jpsi \phi$ control-mode distributions in data. 
In simulation the particle identification (PID) variables for each charged track are resampled as a function of its \ptot, $\pt$ and the number of tracks in the event using $\Lc\to\proton\Km\pip$  
 and $D^{\ast +}\to D^0(\to K^-\pip)\pi^+$ calibration samples from data~\cite{LHCb-DP-2018-001}.  The selection consists of two boosted decision tree (BDT) classifiers. The first classifier,  $\text{BDT}_{\text{sel}}$, is a selection trained on $\Bs \to \jpsi \phi$ simulation and sideband data with the $\jpsi \proton\antiproton$ invariant mass above $5450\mev$ using the $\ptot$, $\pt$, and $\chi^2_{\rm IP}$ variables of the $\Bs$ candidate, the $\chi^2$ probability from the kinematic fit of the candidate, and the impact parameter distances of the two muons. 
 The second classifier, $\text{BDT}_{\text{PID}}$, is trained on $\Bs\to \jpsi p \antiproton$ simulation and sideband data using proton identification variables: the hadron PID from the ring-imaging Cherenkov detectors, the $\ptot$, $\pt$ and $\chi^2_{\rm IP}$ of the protons. The $\text{BDT}_{\text{PID}}$ output selection criterion is chosen by maximising the figure of merit $\mathcal{S}^2/(\mathcal{S}+\mathcal{B})^{3/2}$, where $\mathcal{S}$ and $\mathcal{B}$ are the signal and background yields in a region of $\pm10\mev$ around the \Bs mass peak. These are determined from a fit to the $\jpsi p \antiproton$ invariant-mass distribution in data after the $\text{BDT}_{\text{sel}}$ selection, multiplied by the efficiency of the $\text{BDT}_{\text{PID}}$ output requirement, obtained from simulation and from sideband data, respectively. 
 
 \begin{figure}[t]
  \centering
  \includegraphics[width=0.5\textwidth]{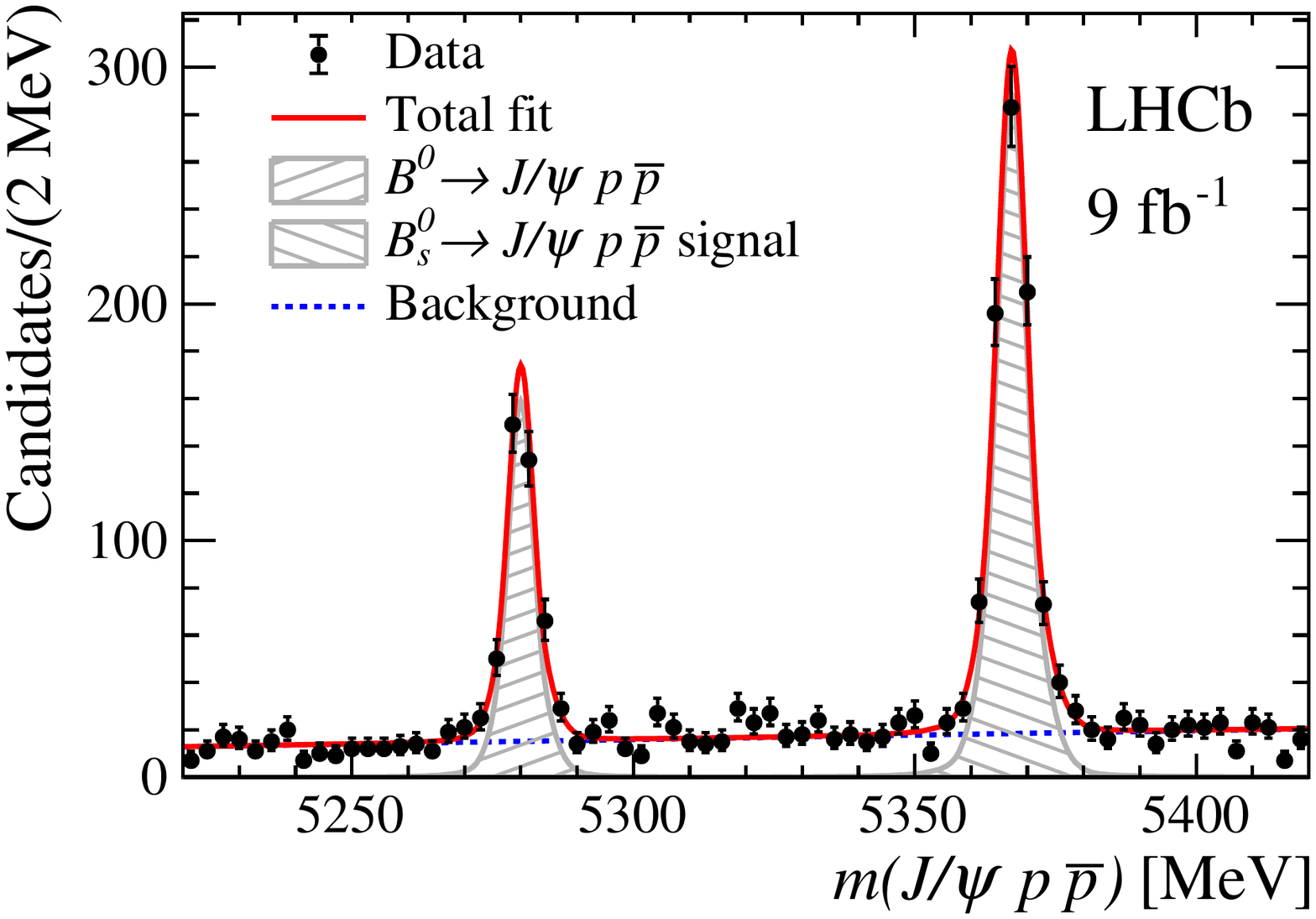}
  \caption{\small Invariant-mass distribution $m(\jpsi p \bar{p})$ for reconstructed signal candidates; the result of the fit described in the text is overlaid. }
  \label{fig:fit_tot}
\end{figure}

\begin{figure}[t]
  \centering
  \includegraphics[width=0.49\linewidth]{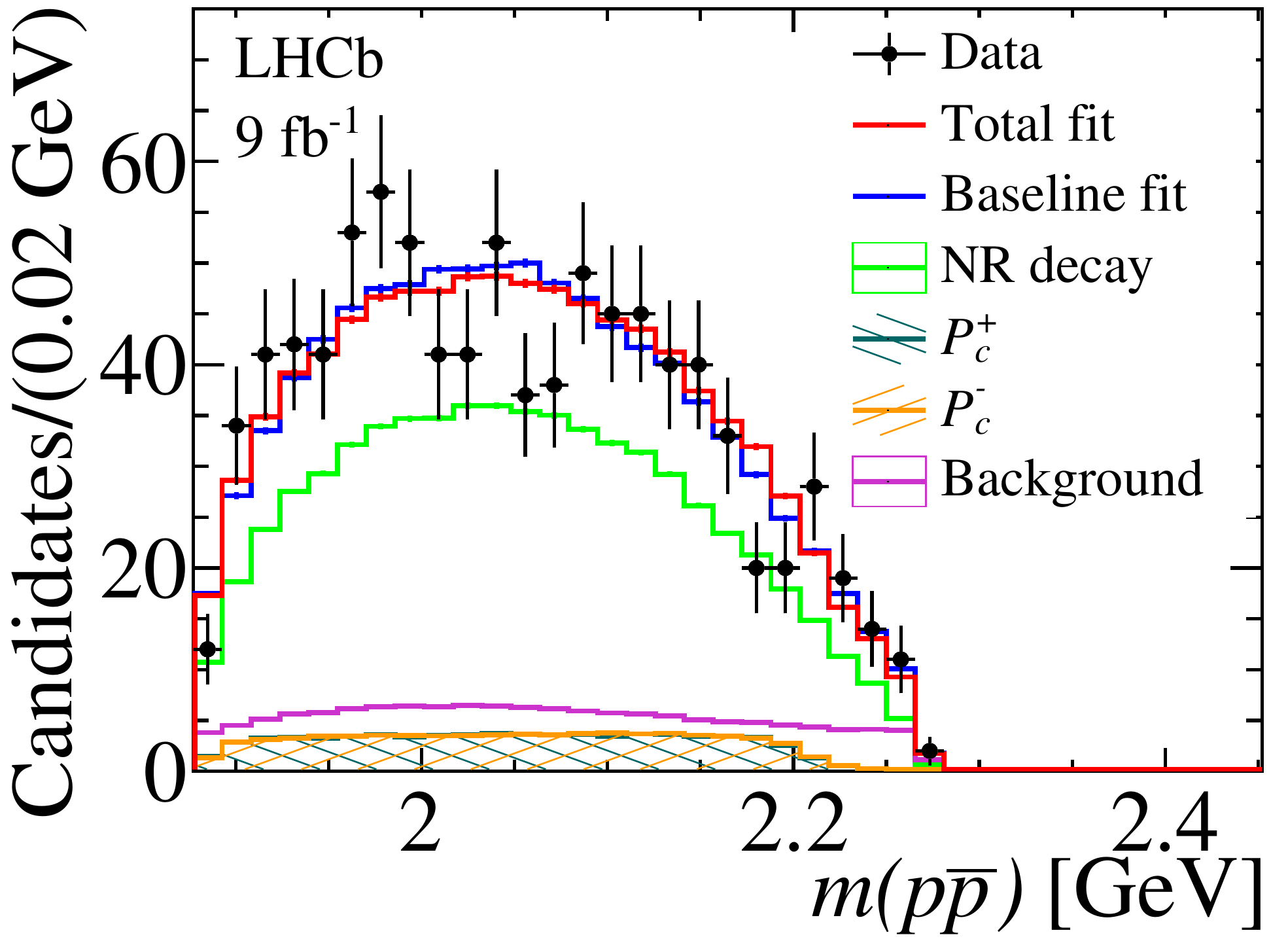}
  \includegraphics[width=0.49\linewidth]{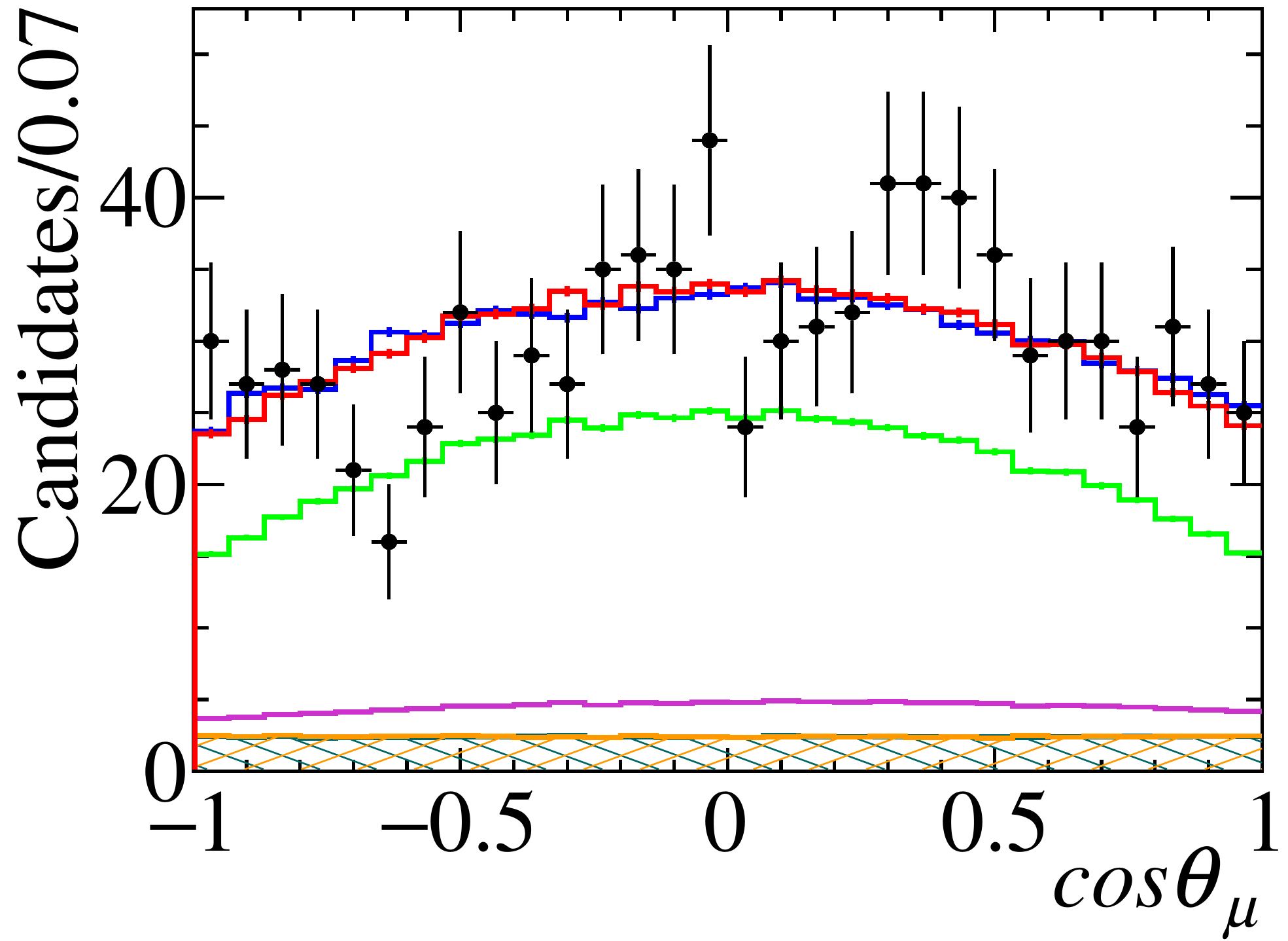}\\ \vspace{1mm}
  \includegraphics[width=0.49\linewidth]{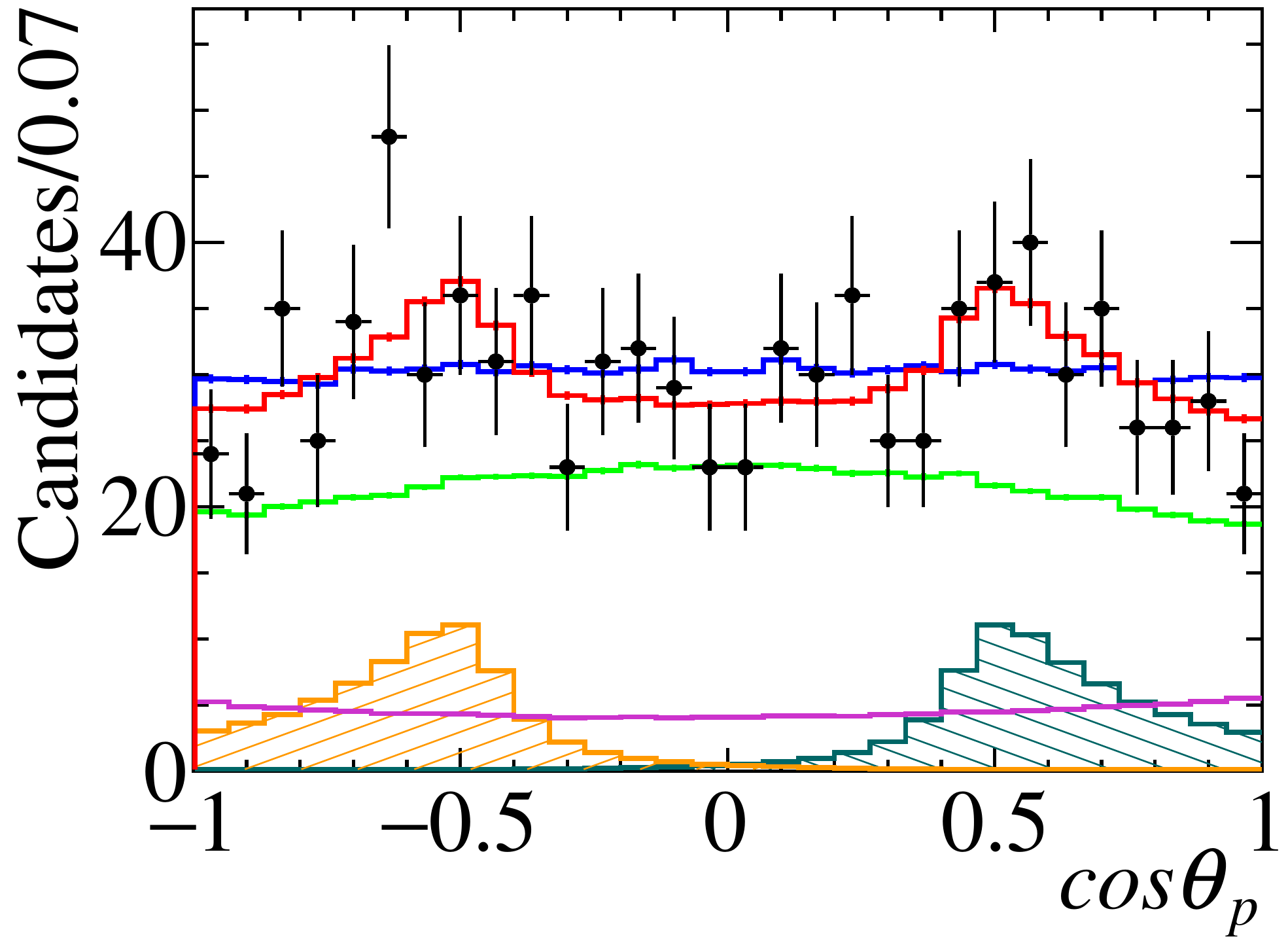}
  \includegraphics[width=0.49\linewidth]{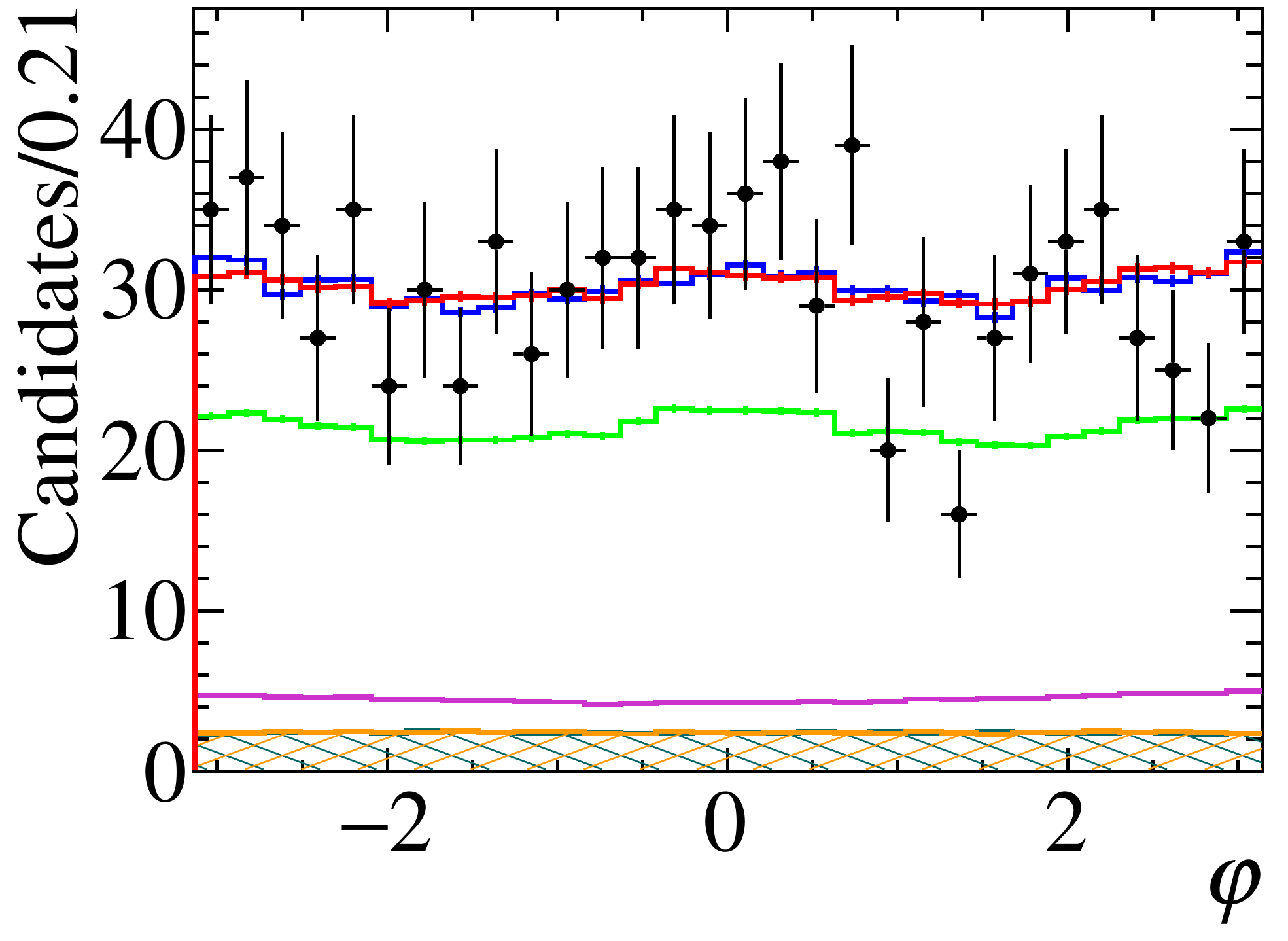}\\
  \vspace{1mm}
  \includegraphics[width=0.49\linewidth]{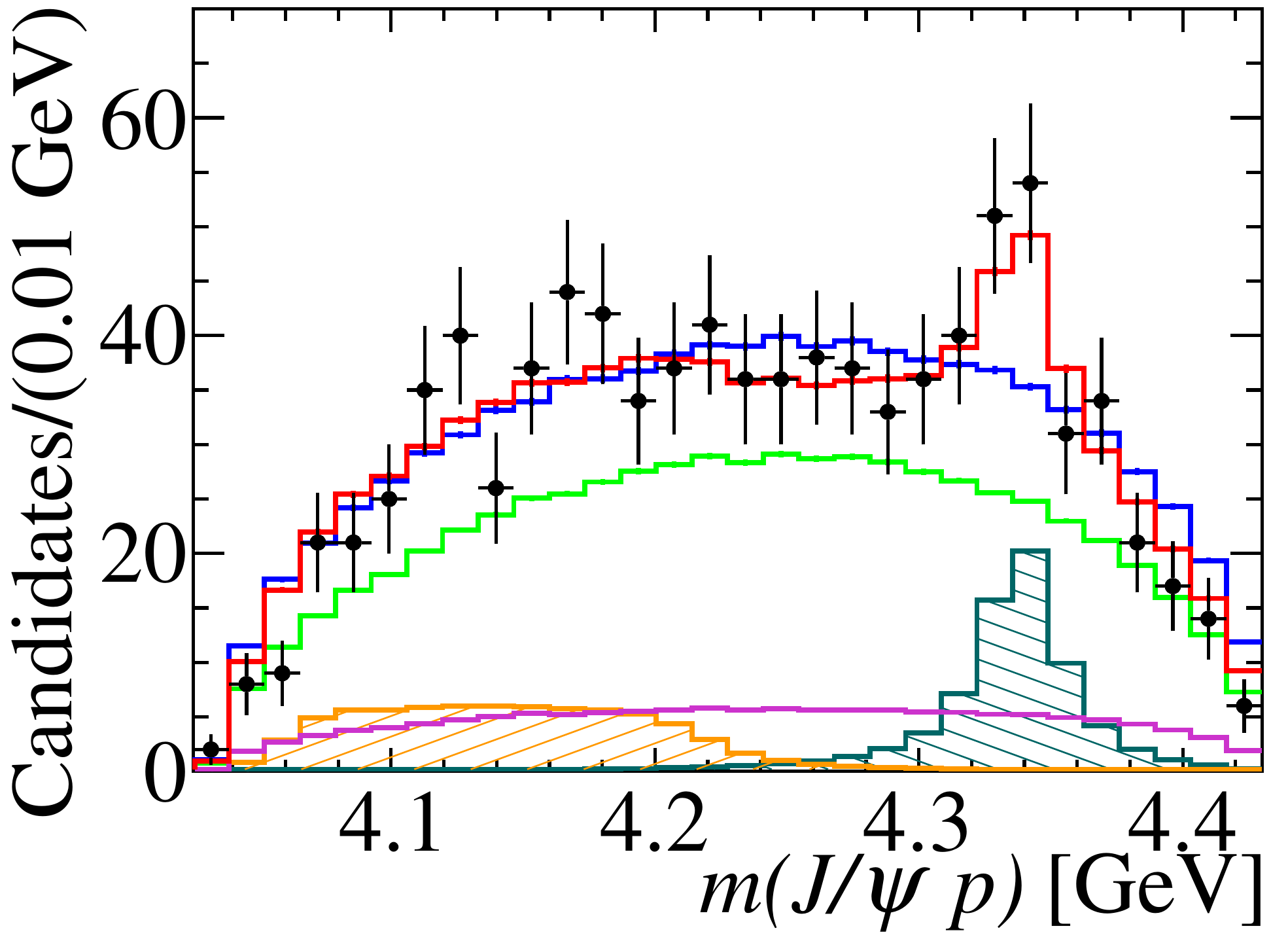}
  \includegraphics[width=0.49\linewidth]{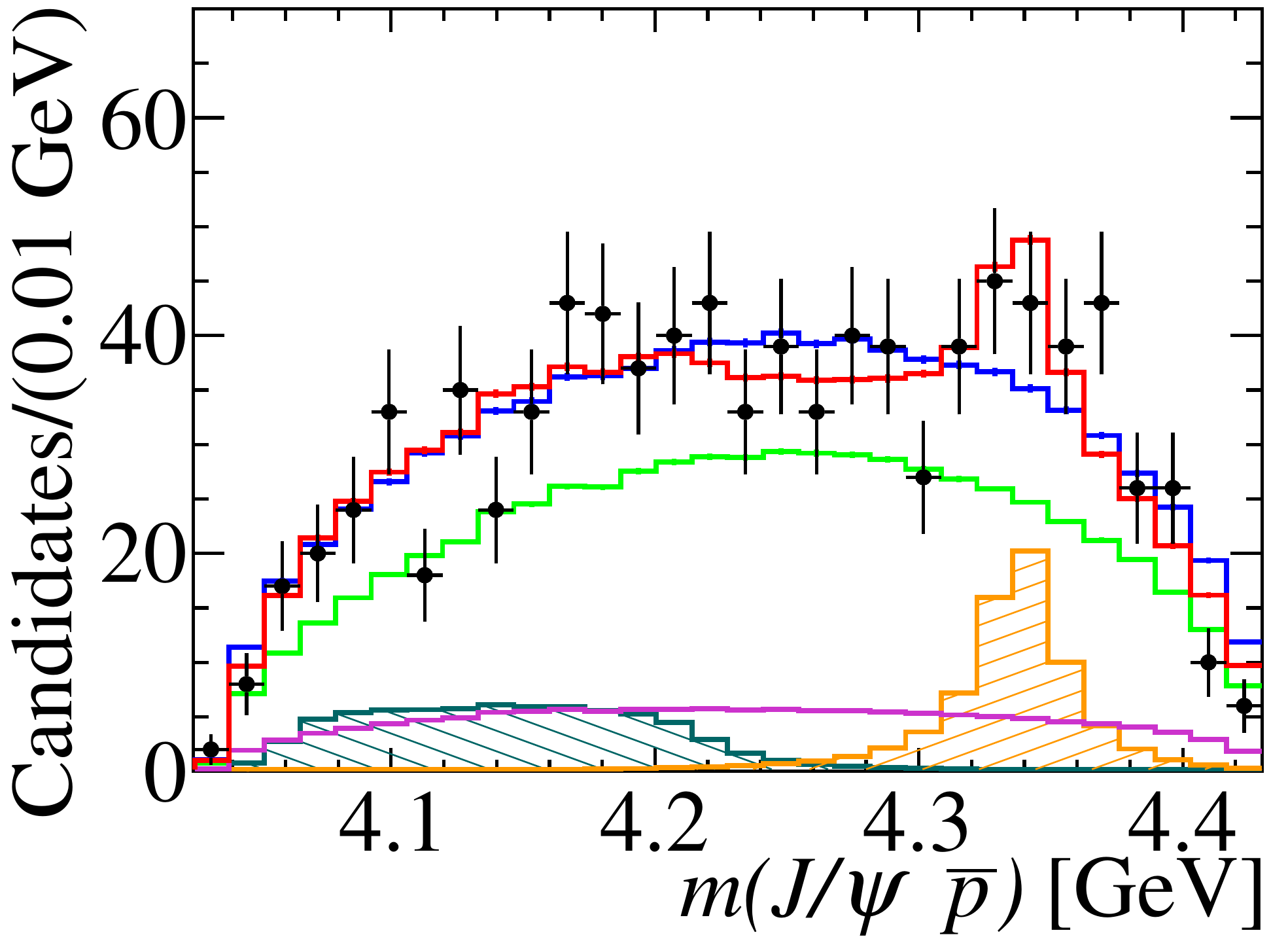}
  \caption{\small One-dimensional projections of the angular $(\cos\theta_{\mu}, \cos\theta_p, \varphi)$ and invariant-mass distributions $(m(\proton\antiproton), m(\jpsi\proton), m(\jpsi\antiproton))$, superimposed with the results of the fit from the baseline model (blue) and the default model (red) comprising a NR term and the $P_c$ contribution.} 
  \label{fig:alt_fit_1+}
\end{figure}

After applying these selection criteria, a maximum-likelihood fit is performed to the $\jpsi p \bar{p}$ invariant-mass distribution, shown in Fig.~\ref{fig:fit_tot}, yielding $797\pm31$ \Bs signal decays. 
The \Bs signal shape is modelled as the sum of two Crystal Ball~\cite{Skwarnicki:1986xj} functions sharing a common peak position, with asymmetric tails describing radiative and misreconstruction effects. The signal-model parameters are determined from simulation and only the \Bs peak position is allowed to vary in the fit to data. 
The combinatorial background is modelled by a first-order polynomial with parameters determined from the fit to data.  The \BJpp component has the same shape as the \Bs signal.  The combinatorial-background fraction in the \Bs signal window of 3$\sigma$ around the mass peak ($[5357, 5378] \mev$) is estimated to be $(14.9\pm0.6)\%$, where $\sigma\approx 3.5 \mev$ is the resolution of the reconstructed invariant mass. 
The $m(\jpsi\proton)$ and $m(\jpsi\antiproton)$ invariant mass distributions of the reconstructed \Bs candidates in the \Bs signal region are shown in the bottom row of Fig.~\ref{fig:alt_fit_1+} (black dots), where hints of structure in the region around (4.3 - 4.4)\gev are present. This Letter investigates the nature of these enhancements, which are not compatible with the pure phase-space hypothesis.

An amplitude analysis of the \Bs candidates is performed under the assumption of \CP symmetry conservation, \ie the dynamics is the same in \Bs and \Bsb decays.  Three interfering decay sequences are considered in the amplitude model: $\Bs \to \jpsi X(\to \proton \antiproton)$, $\Bs \to P_c^+ (\to \jpsi \proton) \antiproton$ and $\Bs \to P_c^-( \to\jpsi\antiproton) \proton$, all followed by a $\jpsi \to \mumu$ decay. These sequences are labelled as the $X$, $P_c^+$ and $P_c^-$ chains, respectively. Since the data sample is not flavour tagged, the distribution of the candidates in the phase space is by construction symmetric for $\jpsi p$ and $\jpsi \bar{\proton}$ final states, and therefore the analysis is sensitive to the sum of possible contributions from $P_c^+$ and $P_c^-$ pentaquark candidates, denoted as $P_c$ in the following. Due to the small sample size and since the \Bs or \Bsbar flavour is not identified, there is no sensitivity to different couplings for the $P_c^+$ and $P_c^-$ states, which are constrained to be equal, up to a phase difference. The amplitude model is based on the helicity formalism of Refs.~\cite{Chung:186421,Jackob1959:at}, which defines a consistent framework for propagating spin correlations through relativistic decay chains. To align the spin of the different decay chains, the prescription in Ref.~\cite{DPdeco} is followed. Details about the amplitude definition are given in the Supplemental material. 

Candidates in the \Bs signal region are used to perform an amplitude fit in the four-dimensional phase space, $(m_{\proton\antiproton}, \vec{\Omega})$. This phase space is defined by the invariant mass $m_{\proton\antiproton}$ of the $\proton \antiproton$ pair and $\vec{\Omega}=(\theta_p, \theta_{\mu}, \varphi)$, where $\theta_p, \theta_{\mu}$ are the two helicity angles of the $\proton$ and the $\mu^-$ in the $X$ and $\jpsi$ rest frame, respectively, and $\varphi$ is the azimuthal angle between the decay planes, of the $\mu^-\mu^+$ and the $\proton\antiproton$ pairs. The distributions of $(m_{\proton\antiproton}, \cos\theta_{\mu}, \cos\theta_p,  \varphi)$, together with the $m(\jpsi\proton)$ and $m(\jpsi\antiproton)$ invariant mass projections, are shown in Fig.~\ref{fig:alt_fit_1+} for selected candidates. 

The amplitude fit minimises the negative log-likelihood function
\begin{align}
-2 \log \mathcal{L}(\vec{\omega})=-2  \sum_{i}\log [ (1-\beta)\mathcal{P}_{\text{sig}}\left(m_{p \bar{p}, i}, \Omega_{i} |\vec{\omega}\right) \nonumber \\ + \beta \mathcal{P}_{\text{bkg}} \left(m_{p \bar{p}, i}, \Omega_{i}\right) ],
\end{align}
where the total probability density function (\PDF) calculated for $i^{\text{th}}$ candidate has a signal, $\mathcal{P}_{\text{sig}}$, and a background, $\mathcal{P}_{\text{bkg}}$, component where $\beta$ is the fraction of background events observed within the \Bs signal window. 
The signal \PDF is proportional to the matrix element squared, $|\mathcal{M}\left(m_{p \bar{p}, i}, \Omega_i | \vec{\omega}\right)|^2$, and depends on the fit parameters, $\vec{\omega}$, \ie the couplings, the masses and the widths, which define the contributing resonances
\begin{align}
    \mathcal{P}_{\mathrm{sig}}\left(m_{p \bar{p}, i}, \Omega_i | \vec{\omega}\right) \equiv \nonumber \\ \frac{1}{I(\vec{w})}\left|\mathcal{M}\left(m_{p \bar{p} i}, \Omega_i | \vec{\omega}\right)\right|^{2} \Phi\left(m_{p \antiproton, i}\right) \epsilon\left(m_{p \bar{p}, i}, \Omega_i\right). \label{eq:Psig}
\end{align}
The phase-space element is $\Phi\left(m_{p \antiproton, i}\right)=|\vec{p}||\vec{q}|$, where $\vec{p}$ is the momentum of the $X$ system in the $\Bs$ rest frame and $\vec{q}$ is the proton momentum in the $X$ rest frame.
The efficiency, $\epsilon\left(m_{p \bar{p}, i}, \Omega_i\right)$, is included in the \PDF, and is parameterised by a Legendre polynomial expansion on the four-dimensional phase space. 
The denominator, $I(\vec{\omega})$, normalises the probability. The fit fractions of each signal component are defined as the corresponding \PDF integral divided by $I(\vec{\omega})$.
The background contribution, $\mathcal{P}_{\mathrm{bkg}}$, is parameterised by the product of one-dimensional Legendre polynomials describing candidates in the \Bs sideband region of $[5420, 5700]\mev$.

No well established  resonances are expected either in the $\proton\antiproton$ or in the $\jpsi \proton$ and $\jpsi \antiproton$ channels. However, some resonances could potentially decay into $\proton\antiproton$~\cite{PDG20}, 
\eg the $f_J(2220)$\cite{PhysRevLett.76.3502} and the $X(1835)$~\cite{Ablikim:2005um, BESIII:2011aa}; thus they have been included in alternative models.
The simplest model used to fit the data has no resonant contributions in the $P_c^+$, $P_c^-$ and $X$ decay chains, and is denoted as the baseline model. 
This model includes a nonresonant (NR) contribution in the $X$ decay sequence with spin-parity quantum numbers equal to $J^P=1^-$, which has $S$-wave terms in both its production and decay. Indeed, due to the low $Q$-value of the decay, the $S$-wave contribution is expected to be favoured since higher values of orbital momentum are suppressed.
Models including NR contributions with different quantum numbers (\ie $J^P=0^{\pm}, 1^+$) are excluded because their $-2\log\mathcal{L}$ values are significantly worse than that of the $J^P=1^-$ hypothesis.

Due to the limited sample size, the baseline model is described by two independent $LS$ couplings for both $\Bs \to \jpsi X$ and $X\to \proton \bar{\proton}$ decays, where $L$ is the decay orbital angular momentum, and $S$ is the sum of spins of the decay products.  
Fixing the two lowest orbital momentum couplings as the normalisation choice and three parameters, which are consistent with zero, reduces the number of free parameters to three.

The fit results of the baseline model are shown in Fig~\ref{fig:alt_fit_1+}. The baseline model does not describe the data distribution, with a $\chisq$ goodness-of-fit test result of $\chisq/ndf = 64/38$ corresponding to a $p$-value of  ${4\times10^{-5}}$. 
Therefore, two resonant contributions from $P_c^+$ and $P_c^-$ are added, with identical masses, widths and couplings.
First, the $P_c(4312)$ state previously observed by the \lhcb experiment in the $\Lb \rightarrow \jpsi \proton \Km$ analysis~\cite{LHCb-PAPER-2019-014} is included in the model 
with mass and width fixed at their known values. The broad $P_c$ structure with a  mass around $4380\mev$, observed in 2015~\cite{LHCb-PAPER-2015-029}, is not considered in this fit, since the helicity formalism used in Ref.~\cite{LHCb-PAPER-2015-014} requires modifications in order to properly align the half-integer spin particles of different decay chains and, thus, those results need to be confirmed with an updated analysis of $\Lb \rightarrow \jpsi \proton \Km$ data~\cite{marangotto2020helicity, wang2020novel}. 
In this analysis no evidence for the $P_c(4312)$ state is found since the $p$-value, computed from the $-2\Delta\log\mathcal{L}$ of the alternative fit with respect to the default model, is measured to be $0.5$. Exploiting the $\text{CL}_s$ method~\cite{CLs}, an upper limit on the modulus of its coupling is set to $0.043$ at 90\% of confidence level, which corresponds to a fit fraction of $2.86\%$. 
A model with a new $P_c^{\pm}$ state given a free mass and width is chosen as the default model. Different spin-parity hypotheses for the $P_c$ states are investigated, \ie $J^P=1/2^{\pm}$ and $J^P=3/2^{\pm}$. Due to a limited sample size, only the lowest values of $L$ are considered and the same coupling is assumed for all $J^P$ hypotheses, resulting in two free parameters: the modulus $A(P_c)$ and the phase $\phi(P_c)$ of the coupling. The seven fit parameters $\vec{\omega}$ contain the baseline model parameters, see Eq.~\ref{eq:Psig}, the coupling $[A(P_c),\phi(P_c)]$, the mass and width of the $P_c$ state.

The fit result for the $J^P=1/2^+$ hypothesis of the $P_c^+$ state is shown in Fig.~\ref{fig:alt_fit_1+}. The $\chi^2/ndf$ is $36.7/36.8$, where the number of degrees of freedom $ndf$ is determined from fits to the $\chi^2$ distribution extracted from pseudoexperiments.  The statistical significance is estimated from pseudoexperiments generated with the baseline model and fitted with the default model, using amplitude parameters determined by the fit to  data. The mass and width of the $P_c$ states are not defined in the baseline model, thus multiple fits to the same pseudodata are performed to account for the look-elsewhere effect, scanning the initial mass value in intervals of size $50\mev$. 
The test statistic $t$ is built as the maximum of the $-2\log\mathcal{L}$ difference between the baseline and the default model~\cite{doi:10.1142/6096} among all the fits obtained by scanning the initial mass values.
 The $p$-value is computed using a frequentist method as the fraction of pseudoexperiments with $t$ larger than the $t_{\text{data}}$ value from the fits to data. The $p$-value ranges between $0.02\%$ and $0.2\%$ for different $J^P$ hypotheses, the lowest being associated to $1/2^+$ and the highest to $3/2^+$, as reported in the Supplemental material. These $p$-values correspond to a signal significance in the range of $3.1$ to $3.7 \sigma$, providing evidence for a new pentaquark-like state. 
Using the $\text{CL}_s$ method~\cite{CLs}, none of the $J^P$ hypotheses considered can be excluded at 95\% confidence level. 

The hypothesis of a glueball state with mass equal to $2230\mev$ and width of around $20\mev$~\cite{Hsiao:2014tda} is also tested, by adding to the default model a resonance in the $X$ decay chain with fixed mass and width. No evidence of $f_J(2220)$ is observed, as the fit with this contribution gives a $p$-value, computed from the $-2\Delta\log\mathcal{L}$ with respect to the default model, of $0.75$ and an associated complex coupling of $[-0.04\pm0.09, -0.06\pm0.16]$.

Systematic uncertainties are evaluated for the mass, width, coupling, and fit fractions of the sum of the $P_c^{\pm}$ contributions. For each source of uncertainty, pseudoexperiments are generated according to the alternative model with the same sample size as in data.  The fit to such pseudoexperiments is performed using the default model. The systematic uncertainties, listed in Table~\ref{tab:sys}, are assigned as the mean of the residual distributions between the fitted and the default parameter results.  
The main contributions are due to different NR models for the $X$ decay chain, alternative $J^P$ hypotheses for the $P_c$ state and possible mismodelling of the efficiency distribution. The systematic uncertainty associated to the NR model is obtained including, in addition to the NR term with $J^P=1^-$ and lowest values of $L$ allowed, a $P$-wave resonant contribution with $J^P=0^-$, modelled with a Breit--Wigner lineshape in order to account for possible resonances, such as the $X(1835)$~\cite{Ablikim:2005um, BESIII:2011aa}, decaying to a $\proton \antiproton$ final state. 
Since none of the $J^P$ hypotheses investigated for the $P_c^{\pm}$ state can be excluded, an additional systematic uncertainty is assigned as the difference between the least and the most significant hypotheses. 
Finally, the uncertainty associated with the efficiency parameterisation is evaluated by summing two contributions. The first is obtained by replacing the default efficiency map with one determined from simulation of different data-taking conditions, and the second by using a parameterisation given by the product of one-dimensional functions of the considered fit variables.
Other systematic uncertainties include alternative parameterisation of the background shape and the uncertainty in the background normalisation, which is varied within its statistical uncertainty. The background is parameterised using data in a sideband region around the \Bs invariant-mass peak with $m(\jpsi \proton 
\antiproton) \in [5300, 5350] \mev$ and $m(\jpsi \proton 
\antiproton) \in [5420, 5460] \mev$, to account for variations of the background as a function of the invariant mass. 
The default value of the hadron radius size for the Blatt--Weisskopf coefficients~\cite{Wu_2014}, equal to $3\gev^{-1}$, is replaced by two alternate values, $1.5\gev^{-1}$ and $5\gev^{-1}$.
Fit biases in the parameters estimation are extracted  from the residual distribution of the generated and fitted parameters of pseudoexperiments based on the default model.
Systematic uncertainties from orbital momentum for the NR, $P_c$ contributions, and invariant-mass resolution are found to be negligible. More details about systematic uncertainties can be found in the Supplemental material. The final significance including systematic uncertainties is equal to 3.1$\sigma$, which is the minimal value among the different sources of systematic uncertainty, as reported in Table~\ref{tab:sys}.

\begin{table}[t]
 \centering
 \caption{Systematic uncertainties associated to the mass $M_{P_c}$ (in \mev), width $\Gamma_{P_c}$ (in \mev), modulus of coupling $A(P_c)$, fit fractions $f(P_c)$ (in \%), $p$-values and associated significance ($\sigma$) of the $P_c^{\pm}$ state.}
\begin{tabular}{lllllll}
\hline
Source & $M_{P_c}$ & $\Gamma_{P_c}$ & $A(P_c)$ &\text{$f(P_c)$  } &$p$ (\%) &$\sigma$\\
\hline 
\text{NR($X$) model} & 0.1 & 1.4 & 0.013& 6.4  &0.003 & 4.2\\
\text{$J^P(P_c)$ assignment} & 2	& 12 & 0.100 &5.5&0.2 & 3.1\\
\text{Efficiency} & 0.2 & 4 & 0.012&0.4 &0.001 &4.4 \\
\text{Background} & 0.1 & 2 & 0.001& 0.7 &0.001 &4.3 \\
\text{Hadron radius} & 0.7 & 4 & 0.034 &1.7&0.02 & 3.7\\
\text{Fit bias} & $^{+0.2}_{-0.1}$ & $^{+5}_{-2}$  & $^{+0.040}_{-0.040}$& -- & --& -- \\
\hline
Total & $2$ & $14$ & $0.11$ & 8.6 & -- & 3.1\\
\hline 
\end{tabular} \label{tab:sys}
\end{table}

The mass and width of this new pentaquark-like state are measured to be
\begin{align}
   M_{P_c} = 4337 \ ^{+7}_{-4} \ ^{+2}_{-2} \mev, \\
\text{$\Gamma$}_{P_c} = 29 \ ^{+26}_{-12} \ ^{+14}_{-14} \mev, \nonumber 
\end{align}
where the first uncertainty is statistical and the second systematic.
The analysis of flavour untagged \Bs decays is not sensitive to the $P_c^+$ and $P_c^-$ contributions separately, therefore a single coupling is determined,  which has modulus $A(P_c) = 0.19 \ ^{+0.19}_{-0.08}   \ ^{+0.11}_{-0.11}$ and phase $\phi(P_c)$ consistent with zero, corresponding to a fit fraction of $(22.0 \ ^{+8.5}_{-4.0} \pm 8.6) \%$ for the $P_c$ states. Due to the limited sample size, it is not possible to distinguish among different $J^P$ quantum numbers.  A state compatible with this $P_c$ state is predicted in Ref.~\cite{Shen:2017ayv} with $J^P=1/2^+$. 

In conclusion, an amplitude analysis of $\Bs \to \jpsi \proton\antiproton$ decays is presented, using data collected with the \lhcb detector between 2011 and 2018, and corresponding to an integrated luminosity of 9 \invfb.  No evidence is seen for either a $P_c$ state at a mass of $4312\mev$~\cite{LHCb-PAPER-2019-014} or the glueball state $f_J(2220)$ predicted in Ref. \cite{Hsiao:2014tda}. Unlike in other \B decays~\cite{Abe:2002tw, PhysRevLett.92.131801, PhysRevD.72.051101, LHCb-PAPER-2012-047}, no threshold enhancement is observed in the $\proton \antiproton$ invariant-mass spectrum, which is well modelled by a nonresonant contribution. Evidence for a Breit--Wigner shaped resonance in the $\jpsi \proton$ and $\jpsi\antiproton$ invariant masses is obtained with a statistical significance in the range of 3.1 to $3.7 \sigma$, depending on the assigned $J^P$ hypothesis.

\section*{Acknowledgements}
%
%
\noindent We express our gratitude to our colleagues in the CERN
accelerator departments for the excellent performance of the LHC. We
thank the technical and administrative staff at the LHCb
institutes.
We acknowledge support from CERN and from the national agencies:
CAPES, CNPq, FAPERJ and FINEP (Brazil); 
MOST and NSFC (China); 
CNRS/IN2P3 (France); 
BMBF, DFG and MPG (Germany); 
INFN (Italy); 
NWO (Netherlands); 
MNiSW and NCN (Poland); 
MEN/IFA (Romania); 
MSHE (Russia); 
MICINN (Spain); 
SNSF and SER (Switzerland); 
NASU (Ukraine); 
STFC (United Kingdom); 
DOE NP and NSF (USA).
We acknowledge the computing resources that are provided by CERN, IN2P3
(France), KIT and DESY (Germany), INFN (Italy), SURF (Netherlands),
PIC (Spain), GridPP (United Kingdom), RRCKI and Yandex
LLC (Russia), CSCS (Switzerland), IFIN-HH (Romania), CBPF (Brazil),
PL-GRID (Poland) and NERSC (USA).
We are indebted to the communities behind the multiple open-source
software packages on which we depend.
Individual groups or members have received support from
ARC and ARDC (Australia);
AvH Foundation (Germany);
EPLANET, Marie Sk\l{}odowska-Curie Actions and ERC (European Union);
A*MIDEX, ANR, IPhU and Labex P2IO, and R\'{e}gion Auvergne-Rh\^{o}ne-Alpes (France);  Fondazione Fratelli Confalonieri (Italy);
Key Research Program of Frontier Sciences of CAS, CAS PIFI, CAS CCEPP, 
Fundamental Research Funds for the Central Universities, 
and Sci. \& Tech. Program of Guangzhou (China);
RFBR, RSF and Yandex LLC (Russia);
GVA, XuntaGal and GENCAT (Spain);
the Leverhulme Trust, the Royal Society
 and UKRI (United Kingdom).

\newpage
{\noindent\normalfont\bfseries\Large Supplemental material}
\vspace{1cm}

{\noindent\normalfont\bfseries\large A. Matrix element model}
\vspace{0.5cm}

The amplitude model is constructed using the helicity formalism \cite{Chung:186421} where a two body decay $A \to (1)(2)$ contributes to the amplitude with a term
\begin{equation}
    \widetilde{\mathcal{H}}^{A\to(1)(2)}_{\lambda_1, \lambda_2} \mathcal{D}^{J_A}_{\lambda_A, \lambda_1-\lambda_2} (\phi_1, \theta_1, 0) R(m_{12}).
    \label{eq:2body_ampl}
\end{equation}
where $\lambda$ is the particle helicity, defined as the projection of the spin $J_A$ onto the momentum direction, and $\widetilde{H}^{A\to(1)(2)}$ are complex helicity couplings which describe the decay dynamics and take into account the Jacob-Wick phase factor of particle $(2)$~\cite{Jackob1959:at}\footnote{Particle $(2)$ is the particle with opposite momentum with respect to particle $(1)$ in the particle-$A$ rest frame: $\vec{p}_2^{\{A\}}=-\vec{p}_1^{\{A\}}$. }  
and are defined as $\widetilde{H}^{A\to(1)(2)}_{\lambda_1, \lambda_2} \equiv (-1)^{J_2-\lambda_2}{H}^{A\to(1)(2)}_{\lambda_1, \lambda_2}$.
The Wigner $D$-matrix rotates the initial coordinate system of particle $A$, with the $z$ axis aligned along the helicity axis of $A$, to the coordinate system with the $z$ axis aligned along the particle $1$ helicity axis. The angles $\phi_1$ and $\theta_1$ (known as ``helicity angle" of $A$) represent the azimuthal and polar angles of particle $1$ in the rest frame of $A$. The third angle is set to zero by convention. The last term of Eq. \ref{eq:2body_ampl}, $R(m_{12})$, is the lineshape dependence that contains either the Blatt--Weisskopf coefficients and threshold factors or, if $A$ is a resonant contribution, a Breit--Wigner lineshape. Details are given below. 

The helicity couplings are expressed in the $LS$ basis, where $L$ is the orbital and $S$ is the spin angular momentum, using Clebsch--Gordan coefficients, $B_{L, S}$, as
\begin{align}
    \mathcal{H}_{\lambda_{B}, \lambda_{C}}^{A \rightarrow B C}=\sum_{L} \sum_{S} \sqrt{\frac{2 L+1}{2 J_{A}+1}} B_{L, S}  &\braket{{J_{B}}, {\lambda_{B}}, {J_{C}},   {-\lambda_{C}}}{ {S}, {\lambda_{B}-\lambda_{C}}} \nonumber\\ \times &\braket{{L}, {0}, {S},{\lambda_{B}-\lambda_{C}}}{ {J_A}, {\lambda_B-\lambda_C}}. 
\end{align}
For strong decays, possible $L$ values are constrained by the conservation of parity: ${P_A = P_1 P_2(-1)^L}$.

The four-body phase space of the \BsJpp decay is described by the $\proton \antiproton$ invariant mass, $m(\proton \antiproton)$, the two helicity angles $\theta_p$, $\theta_{\mu}$ of the \proton and \muon in their parent reference frame and the azimuthal angle $\varphi$ between the dihadron and dilepton decay planes. 

In order to align different decay sequences, the cyclic ordering of the final particles is adopted to define the helicity angles, as suggested in Ref. \cite{DPdeco}, and the Jacob-Wick convention for particle (2). The ordering is important in order to guarantee a proper spin matching of the final particles and to ensure that all resonances share the same $y$ axis instead of the opposite axis. For the decay under study, the following ordering is considered: $\Bs \to \jpsi(1) \proton(2) \antiproton(3)$, where the angles are defined with respect to particle (1) for the $P_c^+$ chain, to particle (2) for the $X$ chain and to particle (3) for the $P_c^-$ chain. In particular, in the $X$ chain the \proton direction defines the $X$ helicity angle, while in the $P_c^-\to \bar{p} \jpsi$ chain the $\bar{p}$ direction defines the $P_c^-$ helicity angle. 
The Jacob-Wick phase-factor is also needed in order to properly align the final spin. Indeed, the rotation in the $X\to\proton\antiproton$ decay aligns the spin axis along the \proton momentum, while the rotation in the $P_c^+\to\jpsi p$ frame aligns the spin axis in the direction opposite to the \proton momentum.  Therefore, an additional rotation to align the $z$ axis between the $P_c^+$ and $X$ chains generates the particle (2) phase factor equal to $(-1)^{J_p-\lambda_{p}}$ in the amplitude of the $P_c^+$ chain, where $J_p$ and $\lambda_{p}$ are the spin and the helicity of the proton in the $P_c^+$ rest frame, respectively. 
 
Denoting the $\jpsi$ as $\psi$, the matrix element for the $\Bs \to X \jpsi$ chain is
\begin{align}
 \mathcal{M}^{X}_{\lambda_p, \lambda_{\antiproton}, \Delta\lambda} = \sum_{\lambda_{\psi}, \lambda_{X}}  \widetilde{\mathcal{H}}^{\Bs}_{ \lambda_X,\lambda_{\psi}} \mathcal{R}(m_{\proton \antiproton}^2)  \mathcal{D}^{*J_X}_{\lambda_X, \lambda_p-\lambda_{\antiproton}} (\phi_p^{\{X\}}, \theta_{p}^{\{X\}}, 0) \widetilde{\mathcal{H}}^{X}_{\lambda_p, \lambda_{\antiproton}} \mathcal{D}^{*1}_{\lambda_{\psi}, \Delta\lambda}(\phi_{\mu}^{\{\psi\}}, \theta_{\mu}^{\{\psi\}}, 0), 
\label{eq:pp}
\end{align} 
where the first Wigner-$D$ matrix of the $\Bs$ decay is omitted because the $\Bs$ has spin zero, implying that, for conservation of total angular momentum, the helicity of the $\psi$ is equal to the helicity of $X$ ($|\lambda_X-\lambda_{\psi}| \leq 0$).  The $y$ axis of the $\psi$ rest frame for all three decay sequences is chosen to be parallel to the $y$ axis in the \Bs rest frame to ensure a correct alignment. The angles $\phi_p^{\{X\}}$ and $\theta_p^{\{X\}}$ are the azimuthal and polar angles of the proton momentum in the $X$ helicity frame, while the angles $\theta_{\mu}^{\{\psi\}}$ and $\phi_{\mu}^{\{\psi\}}$ are the polar and azimuthal angles of the $\mu^-$ momentum in the $\psi$ helicity frame. In the \Bs rest frame, the proton momentum projected on the $x$ axis is positively defined in order to satisfy $\phi_p=0$. Since the $\psi$ decay occurs through an electromagnetic interaction, the difference of the muon helicities, $\Delta\lambda$, restricts to the values: $\Delta\lambda = \lambda_{\mun}-\lambda_{\mup}=\pm1$. As the value of $\Delta\lambda=0$ is  suppressed  by $m_{\mu}/m_{\psi}$ for the electromagnetic transition, it is omitted in the summation. The helicity coupling of the $\psi$ decay is, therefore, ignored because it can be absorbed in the \Bs couplings. The factor $\mathcal{R}(m_{\proton \antiproton}^2)$ is the lineshape term of the $\proton \antiproton$ invariant mass, as described below. 

For $J^P(X)=1^-$, which is the best choice to fit the data, there are 3 (2) independent couplings $\widetilde{\mathcal{H}}^{\Bs}_{\lambda_{\psi}, \lambda_X}$ ($\widetilde{\mathcal{H}}^{X}_{\lambda_p, \lambda_{\antiproton}}$) to fit, already reduced by the parity conservation in the strong decay of the $X$ resonance. Those couplings are then expressed in the $LS$ basis; the lowest $L$ and $S$ of the \Bs decay, $B^{\Bs}_{L_{min}S_{min}}$, is always fixed to $(1,0)$ for each contribution. Therefore, the NR contribution contains two complex couplings 
for the $\Bs \to \jpsi X$ decay, associated to relative angular momenta $L_{\Bs}=0,1,2$ and one for the decay $X\to \proton\antiproton$, with $L_{X}=0,2$, due to parity conservation. Here, $L_{\Bs}$ and $L_{X}$ refer to the relative angular momentum between $\jpsi X$ and $p \antiproton$ final states, respectively. 
In addition, due to the limited sample size, the couplings relative to $L_{\Bs}=1$ for the production and to the phase $\phi$ of $L_{X}=2$ are consistent with zero and, therefore, are fixed to zero in the default model, reducing the number of free parameters to three. The number of $LS$ couplings and free parameters are summarised in Table~\ref{tab:param}.

\begin{table}[t]
\centering
\caption{$LS$ couplings and free parameters for the NR $1^{--}$ model and the models with the $P_c$ states for different $J^P$ hypotheses. Here, $(L,S)$ and $(l,s)$ refer to the relative angular momentum and spin of $P_c^+ \antiproton \ (P_c^-\proton)$ and $\jpsi p \ (\jpsi \antiproton)$ final states, respectively, for the $P_c^+$ ($P_c^-$) decay chain. The free parameters are denoted as $(A, \phi)$, where $A$ is the modulus and $\phi$ the phase of the coupling.} \label{tab:param}
\begin{tabular}{c|cc}
\hline
Model & $LS$ couplings & Free parameters \\
\hline
\hline
NR $J^P(X)=1^{--}$ & $(L,S)_{\Bs}$=(0,0),(2,2), $(L,S)_{X}$=(0,1),(2,1) & $(A, \phi)_{L_{\Bs}=2}, (A, 0)_{L_{X}=2}$\\
$J^P(P_c) = 1/2^{+}$ & $(L,S)=(0,0), (l,s)=(1,1/2)$ & $(A, 0)$\\
 $J^P(P_c) =1/2^-$ & $(L,S)=(0,0), (l,s)=(0,1/2)$ & $(A, \phi)$ \\
 $J^P(P_c) =3/2^-$ & $(L,S)=(1,1), (l,s)=(0,3/2)$ & $(A, 0)$ \\
 $J^P(P_c) =3/2^+$ & $(L,S)=(1,1), (l,s)=(1,1/2)$ & $(A, \phi)$ \\
\hline
\end{tabular}
\end{table}

Similarly, the matrix elements for the $P_c^+$ and $P_c^-$ decay chains are given by
\begin{align}
\mathcal{M}^{P_c^+}_{\lambda_p, \lambda_{\antiproton}, \Delta\lambda}= \sum_{\lambda_{\psi}, \lambda_{P_c^+}}   {\widetilde{\mathcal{H}}^{\Bs\to P_c^+ \antiproton}_{\lambda_{P_c^+}, \lambda_{\antiproton}}}
\mathcal{R}(m_{J/\psi p}^2)  \mathcal{D}^{*J_{P_c}}_{\lambda_{P_c^+}, \lambda_{\psi}-\lambda_{\proton}} (\phi_{\psi}^{\{P_c^+\}}, \theta_{\psi}^{\{P_c^+\}}, 0)   {\widetilde{\mathcal{H}}^{P_c^+ \to \psi p}_{\lambda_{\psi}, \lambda_{p}}} \nonumber \\
 \cdot  \mathcal{D}^{*1}_{\lambda_{\psi}, \Delta\lambda} (\phi^{\{P_c^+\}}_{\mu}, \theta_{\mu}^{\{P_c^+\}}, 0)  
\label{eq:Pcp_ampl}
\\\nonumber
\\
\mathcal{M}^{P_c^-}_{\lambda_p, \lambda_{\antiproton}, \Delta\lambda} = \sum_{\lambda_{\psi}, \lambda_{P_c^-}} {\widetilde{\mathcal{H}}^{\Bs\to P_c^- p}_{\lambda_{P_c^-}, \lambda_{p}}}
\mathcal{R}(m_{J/\psi \antiproton}^2) \mathcal{D}^{*J_{P_c}}_{\lambda_{P_c^-}, \lambda_{\antiproton}-\lambda_{\psi}} (\phi_{\antiproton}^{\{P_c^-\}}, \theta_{\antiproton}^{\{P_c^-\}}, 0)   {\widetilde{\mathcal{H}}^{P_c^- \to \antiproton \psi}_{\lambda_{\antiproton},\lambda_{\psi}}} \nonumber \\
\cdot \mathcal{D}^{*1}_{\lambda_{\psi}, \Delta\lambda} (\phi^{\{ P_c^-\}}_{\mu},\theta_{\mu}^{\{P_c^-\}}, 0)  
\label{eq:Pcm_ampl}
\end{align}
where, as above, for angular momentum conservation, the following relations between helicities hold: $\lambda_{P_c^+} = \lambda_{\antiproton}$ and $\lambda_{P_c^-} = \lambda_{\proton}$ for the $P_c^+$ and $P_c^-$ chains, respectively. The $x$-axes are chosen in order to have $\phi_{\psi}^{\{P_c^+\}}=0$ and $\phi_{\antiproton}^{\{P_c^-\}}=0$. The angle $ \theta_{\psi}^{\{P_c^+\}}$ ($\theta_{\antiproton}^{\{P_c^-\}}$) is the polar angle of the $\psi$ (\antiproton) momentum in the $P_c^+$ ($P_c^-$) helicity frame, while $\theta_{\mu}^{\{P_c^{\pm}\}}$ and  $\phi_{\mu}^{\{P_c^{\pm}\}}$ are the polar and azimuthal angles of the $\mu^-$ momentum in the $\psi$ helicity frame in the two decay chains. 
Due to the limited sample size, from the fit to data neither two unique couplings for the $P_c$ states nor a relative phase between them can be extracted. A relation between the helicity couplings of $P_c^+$ and $P_c^-$ is thus imposed 
\begin{align} \label{eq:coupl_rel}
    {\mathcal{H}^{\Bs\to P_c^- p}_{\lambda_{P_c^{-}}, \lambda_{p}}} = -{\mathcal{H}^{\Bs\to P_c^+ \antiproton}_{\lambda_{P_c^{+}}, \lambda_{\antiproton}}} \nonumber \\ 
    {\mathcal{H}^{P_c^- \to \antiproton \psi}_{\lambda_{\antiproton},\lambda_{\psi}}} = (-1)^{J_{Pc}-s_p-s_{\psi}}{\mathcal{H}^{P_c^+ \to \psi p}_{\lambda_{\psi}, \lambda_{p}}}
\end{align}
where the sign in the second line comes from the permutation of the final helicities, while the one of the first line depends on the weak dynamics and is chosen as such.
This choice is arbitrary and does not impact the results because, due to limited sample size, there is no sensitivity to the relative interference between $P_c^{\pm}$ states. It is made such that it simplifies the model by constraining the two $P_{c}^{\pm}$ couplings to be equal and obtains the same interference pattern for all $J^P$ hypotheses, as discussed below in Eq. \ref{eq:CP_rel}. The helicity couplings expressed in the LS basis are reduced to the lowest $L$ values allowed. There is one independent coupling to fit, which is reduced to one free parameter for $J^P=1/2^+$ and $3/2^-$ because the phase $\phi$ of the coupling is consistent with zero due to the limited sample size, as summarised in Table~\ref{tab:param}.  

A single resonant contribution in the $P_c$ decay chains is parametrised by
\begin{equation}
    \mathcal{R}(m) = B_{L}^{\prime} \left(p, p_{0}, d\right) \left(\frac{p}{M_{B_s^{0}}}\right)^{L_{B_{s}^{0}}} \mathrm{BW}\left(m | M_{0}, \Gamma_{0}\right) B_{L_{P_{c j}}}^{\prime}\left(q, q_{0}, d\right)\left(\frac{q}{M_{0}}\right)^{L_{P_{c j}}}. 
\end{equation}
where 
\begin{align}
&\mathrm{BW}\left(m | M_{0}, \Gamma_{0}\right)=\frac{1}{M_{0}^{2}-m^{2}-i M_{0} \Gamma(m)},
 \end{align}
 is the Breit--Wigner function which includes the mass-dependent width
 \begin{align}
&\Gamma(m)=\Gamma_{0}\left(\frac{q}{q_{0}}\right)^{2 L_{P_{c j}}+1} \frac{M_{0}}{m} B_{L_{P_{c j}}}^{\prime}\left(q, q_{0}, d\right)^{2}.
\end{align}
Here, $p$ is the momentum of the $P_c$ resonance in the \Bs rest frame, $q$ is the momentum of one of the decay products in the $P_c$ rest frame, while the momenta $p_0$ and $q_0$ denote their values  at the resonance peak ($m = M_0$), and $d$ is the hadron radius size fixed to $3\gev^{-1}$ in the default fit and varied to $1.5\gev^{-1}$ and $5\gev^{-1}$ in the systematic uncertainty evaluation. The orbital momentum of the \Bs decay is denoted as $L_{B_{s}^{0}}$ and that of the $P_c$ decay as $L_{P_{c j}}$.
In order to properly account for the suppression due to higher values of $L$, the Blatt-Weisskopf coefficients are used together with the orbital barrier factor $p^L B_L(p, p_0, d)$.
 For NR contributions in the $X$ chain,  $\text{BW}(m)$ is set to 1 and $M_{0 (NR)}$ to the midrange mass. 

To sum the amplitudes from different decay chains coherently, the final state helicities of $\proton, \antiproton$ and $\muon$ in the $P_c$ chains must be rotated in order to match the helicities of the $X$ chain
\begin{align}
     \left|{\mathcal{M}}\right| ^2= \sum_{\lambda_p} \sum_{\lambda_{\antiproton}}\sum_{\Delta\lambda} \mid \mathcal{M}^{X} + e^{i \Delta\lambda \cdot \alpha_{\mu}} \sum_{\lambda_p^{P_c}}\sum_{\lambda_{\antiproton}^{P_c}}  d^{1/2}_{\lambda_{\antiproton}^{P_c}, \lambda_{\antiproton}} (-\theta_{\antiproton}^{P_c^+})  d^{1/2}_{\lambda_{p}^{P_c},\lambda_{p}} (\theta_{p}^{P_c^+})\mathcal{M}^{P_c^+}(\lambda_p^{P_c}, \lambda_{\antiproton}^{P_c}, \Delta\lambda)  + \nonumber \\ + e^{i \Delta\lambda \cdot \alpha_{\bar{\mu}}}\sum_{\lambda_{p}^{P_c}} \sum_{\lambda_{\antiproton}^{P_c}}  d^{1/2}_{\lambda_{p}^{P_c}, \lambda_{p} } (\theta_{p}^{P_c^-}) d^{1/2}_{\lambda_{\antiproton}^{P_c},\lambda_{\antiproton}} (-\theta_{\antiproton}^{P_c^-})\mathcal{M}^{P_c^-}(\lambda_{\antiproton}^{P_c},\lambda_{p}^{P_c}, \Delta\lambda) \mid ^2 \label{eq:ampl_Wigner}
\end{align}
where $\theta_{p}^{P_c^{\pm}}$ ( $\theta_{\antiproton}^{P_c^{\pm}}$) are the polar angles in the $\proton$ ($\antiproton$) rest frame between the boost directions from the $P_c^{\pm}$ to the $X$ rest frames and $\alpha_{\mu}$ is the azimuthal angle in the $\psi$ rest frame to correct the muon helicity states in the two chains. 
Regarding the definition of the polar angles, for the $P_c^+$ chain, $\theta_{p}^{P_c^{+}}$ and $\theta_{\bar{\proton}}^{P_c^{+}}$ are defined as the opening angle between the $\bar{p}$ and $\psi$ mesons in the $\proton$ rest frame and between the \proton and the $P_c$ momenta in the $\bar{p}$ rest frame, respectively
\begin{equation}
    \cos\theta_p^{P_c^+} = \hat{p}_{\bar{p}}^{\{p\}} \cdot  \hat{p}_{\psi}^{\{p\}} \ \text{and} \ \cos\theta_{\bar{p}}^{P_c^+} = \hat{p}_{\proton}^{\{\bar{p}\}} \cdot \hat{p}_{P_c}^{\{\bar{p}\}}.
\end{equation}
For the $P_c^-$ chain, the $\theta_{p}^{P_c^{-}}$ and $\theta_{\bar{\proton}}^{P_c^{-}}$ angles are obtained by substituting $\proton \leftrightarrow \antiproton$.
Since the $y$ axis is outgoing from the plane, the rotation to align the $P_c^+$ decay chain to the $X$ decay chain is counterclockwise between the $\jpsi$ and the $\bar{p}$ momenta. While in the $\bar{p}$ rest frame, the rotation from the direction of $P_c^+$ to that of $p$ is clockwise, hence of an angle $-\theta_p^{P_c^+}$. For the $P_c^-$ chain, the rotation is always clockwise in the proton rest frame and counterclockwise in the $\bar{p}$ rest frame. 
For the $\jpsi$ decay, since the muons come from the $\jpsi$ for both decay chains, the polar angle is 0, implying the following: $d^{1/2}_{\lambda_{\mu}^{P_c}, \lambda_{\mu}} = \delta_{\lambda_{\mu}^{P_c}, \lambda_{\mu}}$. However, there is an azimuthal angle $\alpha_{\mu}$ because of the offset in the $x$ axis. 
Since the boost to the $\mu$ rest frame is the same for both decay chains, \textit{i.e.}~always from the $\jpsi$ rest frame, $\alpha_{\mu}$ can be determined in the $\jpsi$ rest frame as
\begin{equation}
    \alpha_{\mu} = \text{atan2}((\hat{z}_3^{\{\psi\} P_c} \times \hat{x}_3^{\{\psi\}P_c}) \cdot \hat{x}_3^{\{\psi\}X},\hat{x}_3^{\{\psi\}P_c}\cdot \hat{x}_3^{\{\psi\}X}),
\end{equation}
where the index $3$ refers to the rest frame after rotations; $\hat{z}_3^{\psi} = \hat{p}_{\mu}^{\psi}$ and the $\hat{x}_3$ axis can be derived as
\begin{align}
    \hat{x}_3^{\{\psi\}P_c} = - a_{z_0 \perp \mu} \\ 
    a_{z_0 \perp \mu} = -\hat{p}_p^{\{\psi\}} + (\hat{p}_{p}^{\{\psi\}} \cdot \hat{p}_{\mu}^{\{\psi\}}) \hat{p}_{\mu}^{\{\psi\}},
\end{align}
as well as
\begin{align}
    \hat{x}_3^{\{\psi\}X} = - a_{z_0 \perp \mu} \\ 
    a_{z_0 \perp \mu} = -\hat{p}_X^{\{\psi\}} + (\hat{p}_{X}^{\{\psi\}} \cdot \hat{p}_{\mu}^{\{\psi}\}) \hat{p}_{\mu}^{\{\psi\}}.
\end{align}
The term aligning the muon helicity states between the two reference frames is given by
\begin{equation}
    \sum_{\lambda_{\mu}} D^{J_{\mu}}_{\lambda_{\mu}^{P_c},\lambda_{\mu}} (\alpha_{\mu}, 0, 0) = \sum_{\lambda_{\mu}} e^{i\lambda_{\mu}^{P_c}\alpha_{\mu}} \delta_{\lambda_{\mu}^{P_c}, \lambda_{\mu}} = e^{i\lambda_{\mu}\alpha_{\mu}}.
\end{equation}
The transformation for the $\mu^-$ is equal to that of the  $\mu^+$ where the azimuthal angle takes a negative sign: $ \alpha_{\mu^{-}} = - \alpha_{\mu^+}$. Considering the transformations of both muons, the final rotation is the multiplication of the two exponentials
\begin{equation}
    e^{i\lambda_{\mu}\alpha_{\mu}} e^{i\lambda_{\bar{\mu}}\alpha_{\bar{\mu}}} = e^{i\alpha_{\mu}(\lambda_{\mu}-\lambda_{\bar{\mu}}) } = e^{i \alpha_{\mu} \Delta\lambda }.
\end{equation}
This implementation is found to be equivalent to the one proposed in Ref. \cite{DPdeco}, where the spin of the $\jpsi$ in the different chains is aligned with a polar rotation before the $\jpsi \to \mu^+\mu^-$ decay. 

Since the flavour of the \Bs candidate is not tagged, the overall amplitude is the average of the \Bs and $\Bsb$ amplitudes, where the $\Bsb$ amplitude is equal to that of \Bs due to the absence of direct \CP violation and is obtained inverting particles with antiparticles, \ie $\proton \leftrightarrow \antiproton$, $\mun \leftrightarrow \mup$ and inverting all azimuthal angles, $\phi \leftrightarrow -\phi$. To impose \CP symmetry, the flavour eigenstates are projected onto the \CP basis of $\B_L$ and $\B_H$.  Enforcing \CP symmetry conservation, the \Bs and $\Bsb$ and their decay products can be written in terms of $CP$ eigenstates. If \CP is applied to the \Bs final states, then
\begin{align}
    \CP \ket{X\jpsi} &= + \ket{X\jpsi}, \nonumber \\
    \CP \ket{P_c^+\antiproton} &= e^{i\phi} \ket{P_c^-\proton},  \nonumber\\
    \CP \ket{P_c^-\proton} &= e^{i\phi} \ket{P_c^+\antiproton},  \label{eq:CP_relation1}
\end{align}
is obtained.
Indeed, the contribution in the $X$ chain is already a \CP eigenstate, and in particular for $J^P=1^-$ a \CP-even one, while the two single $P_c$ are not. However, a combination of the $P_c^+$ and $P_c^-$ contributions can be projected onto a \CP eigenstate orthogonal to the $X$ contribution (\CP-odd) by choosing the specific phase convention: $e^{i\phi} = -1 (-1)^{J_{P_c}-3/2}$, hence resulting in
\begin{equation}
    \CP (\ket{P_c^+}-(-1)^{J_{P_c}-3/2}\ket{P_c^-}) = - (\ket{P_c^+}-(-1)^{J_{P_c}-3/2}\ket{P_c^-})
    \label{eq:CP_rel}
\end{equation}
which spin dependent factor follows from Eq. \ref{eq:coupl_rel}.
Therefore, the interference between the $X$ and the combination of the $P_c$ contributions cancels out upon integration over the Dalitz plane, as they are orthogonal \CP eigenstates.

\vspace{1cm}

{\noindent\normalfont\bfseries\large B. Event-by-event efficiency parameterisation}
\vspace{0.5cm}

\label{sec:eff}
Event-by-event acceptance corrections are applied to the data using an efficiency parameterisation based on the decay kinematics. 
The 4-body phase space of the topology $\Bs\to \jpsi(\to \mup\mun)p \antiproton$ is fully described by four independent kinematic variables, one mass and three angles: $m_{\proton \antiproton}, \theta_{p}, \theta_{\mu}, \varphi$, where the angles are defined as, 
\begin{itemize}
\item  $\theta_{\mu}$ and $\theta_{p}$: the helicity angles defined in the dimuon and dihadron rest frames, respectively;
\item $\varphi$: the azimuthal angle between the two decay planes of the dilepton and dihadron systems.
\end{itemize}
Since the final state is as self-conjugate, the $\proton$ and the $\mun$ are chosen to define the angles, for both $\Bs$ and $\Bsb$. For the signal mode, the overall efficiency, including trigger, detector acceptance and selection procedure, is obtained from simulation as a function of the four kinematic variables, $\vec{\omega} \equiv \{m'_{\proton\antiproton}, \cos\theta_{\mu}, \cos\theta_{p}, \varphi'\}$. Here, $m'_{\proton\antiproton}$ and $\varphi'$ are transformed such that all four variables in $\vec{\omega}$ lie in the range $(-1,1]$. The efficiency is parameterised as the product of Legendre polynomials
\begin{equation}
\label{eqn:eff_param_def}
\varepsilon(\vec{\varphi}) = \sum_{i,j,k,l} c_{i,j,k,l}~P(\cos\theta_{\mu},i)P(\cos\theta_{p},j)P(\varphi',k)P(m'_{\proton\antiproton},l)\nonumber,
\end{equation}
where $P(x,n)$ is a Legendre polynomial of order $n$ in $x\in(-1,1]$. Employing the order of the polynomials as $\{3,7,7,5\}$ for $\{m'_{\proton\antiproton}, \cos\theta_{\mu}, \cos\theta_{p}, \varphi'\}$, respectively, was found to give a good parameterisation. 
The coefficients, $c_{i,j,k,l}$, are determined from the simulation using a moments technique employing the orthogonality of Legendre polynomials
\begin{align}
\label{eqn:eff_coeff_mom}
      c_{i,j,k,l}\; =\; & \frac{C}{\sum w_{n}}\sum_{n=0}^{N_{\rm recon}} w_{n}\left(\frac{2i+1}{2}\right)\left(\frac{2j+1}{2}\right)\left(\frac{2k+1}{2}\right)\left(\frac{2l+1}{2}\right) \\
& \hspace{1cm} \times P(\cos\theta_{\mu},i)P(\cos\theta_p,j)P(\varphi',k)P(m'_{\proton\antiproton},l)
\end{align}
where $w_{n}$ is the per-event weight taking into account both the generator level phase-space element, $\deriv\Phi$, and the kinematic event weights. Simulation samples are employed where $\Bs\to\jpsi\proton\antiproton$ events are generated uniformly in phase space. In order to render the simulation flat also in $m(p\bar{p})$, the inverted phase-space factor, $1/\deriv\Phi$, is considered. The factors of $(2a + 1)/2$ arise from the orthogonality of the Legendre polynomials,
\begin{equation}
\int_{-1}^{+1} P(x, a) P(x, a') \deriv x = \frac{2}{2 a +1}\delta_{ a a'}  ~.
\end{equation}

The sum in Eq. \ref{eqn:eff_coeff_mom} is over the reconstructed events in the simulation sample after all selection criteria. 
The factor $C$ ensures appropriate normalisation and it is computed such that
\begin{align} 
\sum_{n=0}^{N_{\rm gen}} \varepsilon(\vec{x}_n) = N_{\rm rec},
\end{align} 
where $N_{\rm rec}$ is the total number of reconstructed signal events. 

Up to statistical fluctuations, the parameterisation follows the simulated data in all the distributions.

\clearpage
\vspace{1cm}
{\noindent\normalfont\bfseries\large C. Significance studies for different $J^P$ hypotheses of the $P_c$ state}\vspace{0.5cm}

The significance studies for different $J^P$ hypotheses are reported in Table \ref{tab:diff_spin}, together with the fit results. 
The significance is computed using a frequentist approach, by counting the number of pseudoexperiments above the $-2\Delta\log \mathcal{L}$ observed in data. The $p$-value and corresponding two-sided significance are reported in Tab.~\ref{tab:diff_spin}. 

\begin{table}[htb]
\centering
\caption{Results for all spin-parity hypotheses. Values of $-2 \Delta \log\mathcal{L}$ (-2DLL), $p$-value and two-sided significance extracted from pseudoexperiments, together with mass (in \mev), width (in \mev), fit fraction and complex coupling (A,$\phi$) of the $P_c$ states are reported. }
\begin{tabular}{l|llllllll}
\hline
$J^P$      & -2DLL & $p$ ($\times 10^{-3}$) & $\sigma$ & $M_0$ & $\Gamma_0$ & $f(P_c)$(\%)& A$(P_c)$ & $\phi(P_c)$\\
\hline
$1/2^-$ & 26.2 & $0.5\pm0.3$ &3.5 $\pm$ 0.1  & 4335$^{+3}_{-3}$ & 23$^{+11}_{-8}$ & 17.4$^{+7.0}_{-3.8}$  & 0.15$^{+0.07}_{-0.05}$  & 2.8$^{+1.3}_{-1.4}$\\
$1/2^+$ &  26.8    &$0.2\pm0.1$ & 3.7 $\pm$ 0.1  & 4337$^{+7}_{-4}$ & 29$^{+26}_{-12}$ & 22.0$^{+8.5}_{-4.4}$ & 0.19$^{+0.19}_{-0.08}$ & -0.6$^{+2.4}_{-3.0}$ \\
$3/2^-$ &   25.8  &$0.3\pm0.2$& 3.6 $\pm$ 0.1 & 4337$^{+5}_{-3}$ & 23$^{+16}_{-9}$ & 18.6$^{+6.9}_{-3.0}$ & 0.14$^{+0.08}_{-0.05}$ & -1.3$^{+1.9}_{-2.0}$ \\
$3/2^+$ &  23.6  &$2 \pm 1$& 3.1 $\pm$ 0.1 & 4336$^{+3}_{-2}$ & 15$^{+9}_{-6}$ & 11.7$^{+4.2}_{-2.7}$ & 0.10$^{+0.05}_{-0.03}$ & -3.1$^{+0.6}_{-0.6}$\\
\hline
\end{tabular}\label{tab:diff_spin}
\end{table}

\vspace{1cm}
{\noindent\normalfont\bfseries\large D. Systematic uncertainties studies}\vspace{0.5cm}

In this section, more details about systematic uncertainties are presented. The systematic uncertainties are studied using pseudoexperiments generated according to the alternative model with the same statistics as in data. Parameters are then fitted with the default model. For each parameter, the uncertainty is taken as the mean value of the residual distribution between the fitted ensemble and the generated one. The main systematic uncertainties are described in the  document. To estimate the uncertainty associated to the NR model parameterisation for the $X$ decay chain, alternative NR models are used, also with different quantum numbers ($J^P=0^{\pm}, 1^{+}$). Combinations of NR and a resonant term are used to account for possible resonances decaying to a $p\bar{p}$ final state. The model, which yields comparable results in terms of $-\log\cal{L}$, comprises a NR term with $J^P=1^+$ and a Breit--Wigner resonance with $J^P=0^-$, and is used to assign the systematic uncertainty. The impact of this model on the significance of the $P_c$ state is also studied and is found to be negligible. The other alternative models have been excluded based on the  -2$\Delta\log\mathcal{L}$, with respect to the baseline model, which is greater than $40$. 
In addition, amplitude models including all values of orbital momentum for both the $X$ and the $P_c$ decay chains are considered and their effect is found to be negligible. 
For the systematic uncertainty associated with the efficiency parameterisation, two different contributions are summed in quadrature. First, the default efficiency map extracted from all simulation events is replaced by using simulations from the Run~1 and the Run~2 data-taking periods, separately. Second, a simplified parameterisation with one-dimensional Legendre polynomials is considered, neglecting the effect of correlations among the phase-space variables. Other sources of uncertainty regard different $J^P$ assignment for the $P_c$ state, the background parameterisation, the hadron radius size and the fit bias. The last uncertainty is estimated from pseudoexperiments by generating and fitting the parameters of the default model. Possible biases on the fit parameters are accounted for as the mean of the residual distributions between the generated and the fitted values.   
\vspace{1cm}

{\noindent\normalfont\bfseries\large E. Dalitz plot distribution}
\vspace{0.5cm}

\begin{figure}[t]
    \centering
    \includegraphics[width=0.45\textwidth]{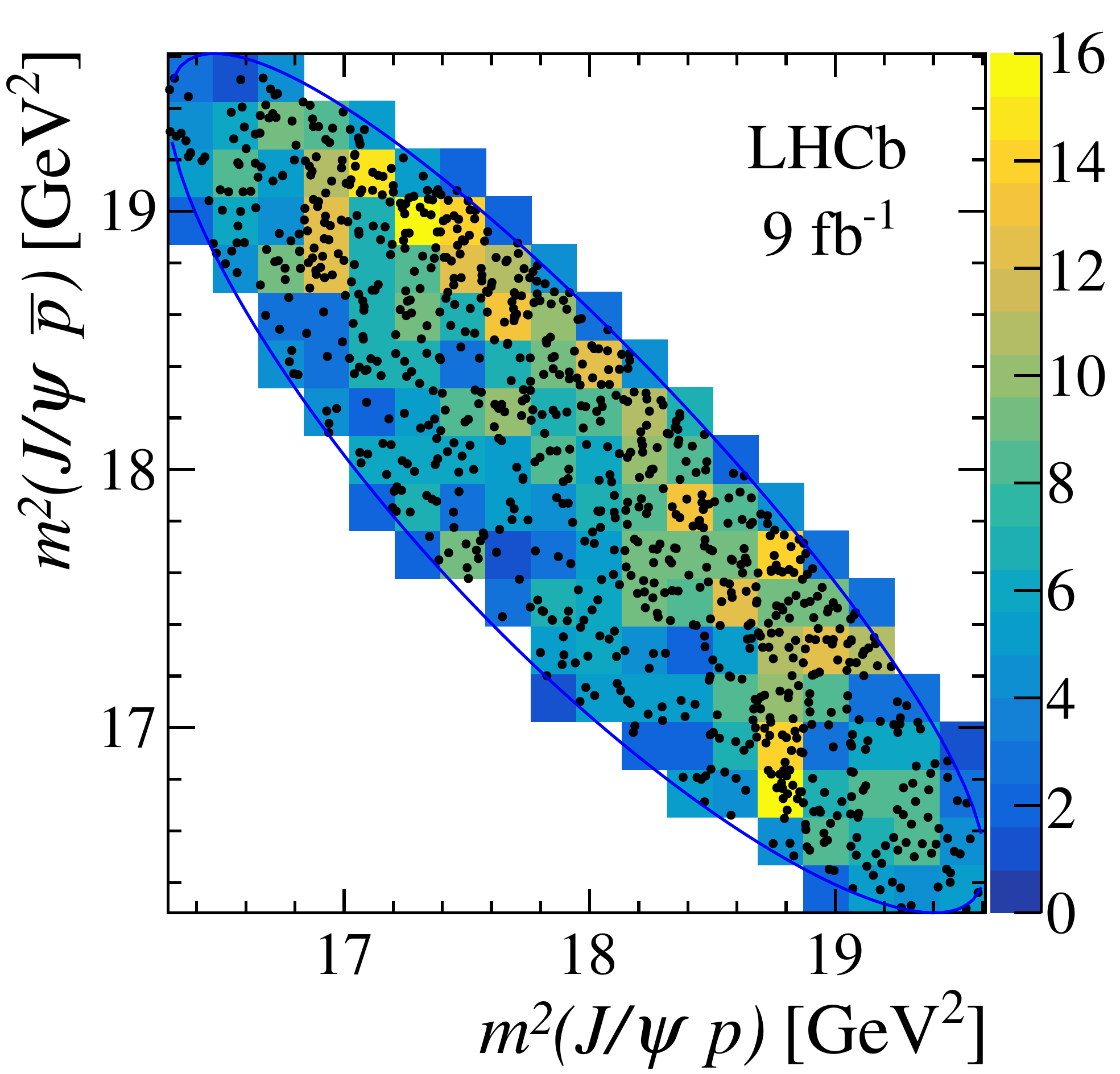}
    \caption{Dalitz plot distribution for reconstructed candidates (black dots)  within the \Bs signal region. The colour scale represents the number of candidates in each Dalitz plot interval.}
    \label{fig:Dalitz}
\end{figure}

The Dalitz plot distribution of the reconstructed \Bs candidates in the \Bs signal region is shown in Fig.~\ref{fig:Dalitz}, where hints of horizontal and vertical bands in the region around $(18.8-19.0)\gev^2$ are present in the $m^2(\jpsi\proton)$ and $m^2(\jpsi\antiproton)$ distributions, respectively. 

\vspace{1cm}

{\noindent\normalfont\bfseries\large F. Maximum and minimum of $m(\jpsi \proton)$ and $m(\jpsi \antiproton)$ distributions}\vspace{0.5cm}

\begin{figure}[hpt]
\centering
\includegraphics[width=0.45\textwidth]{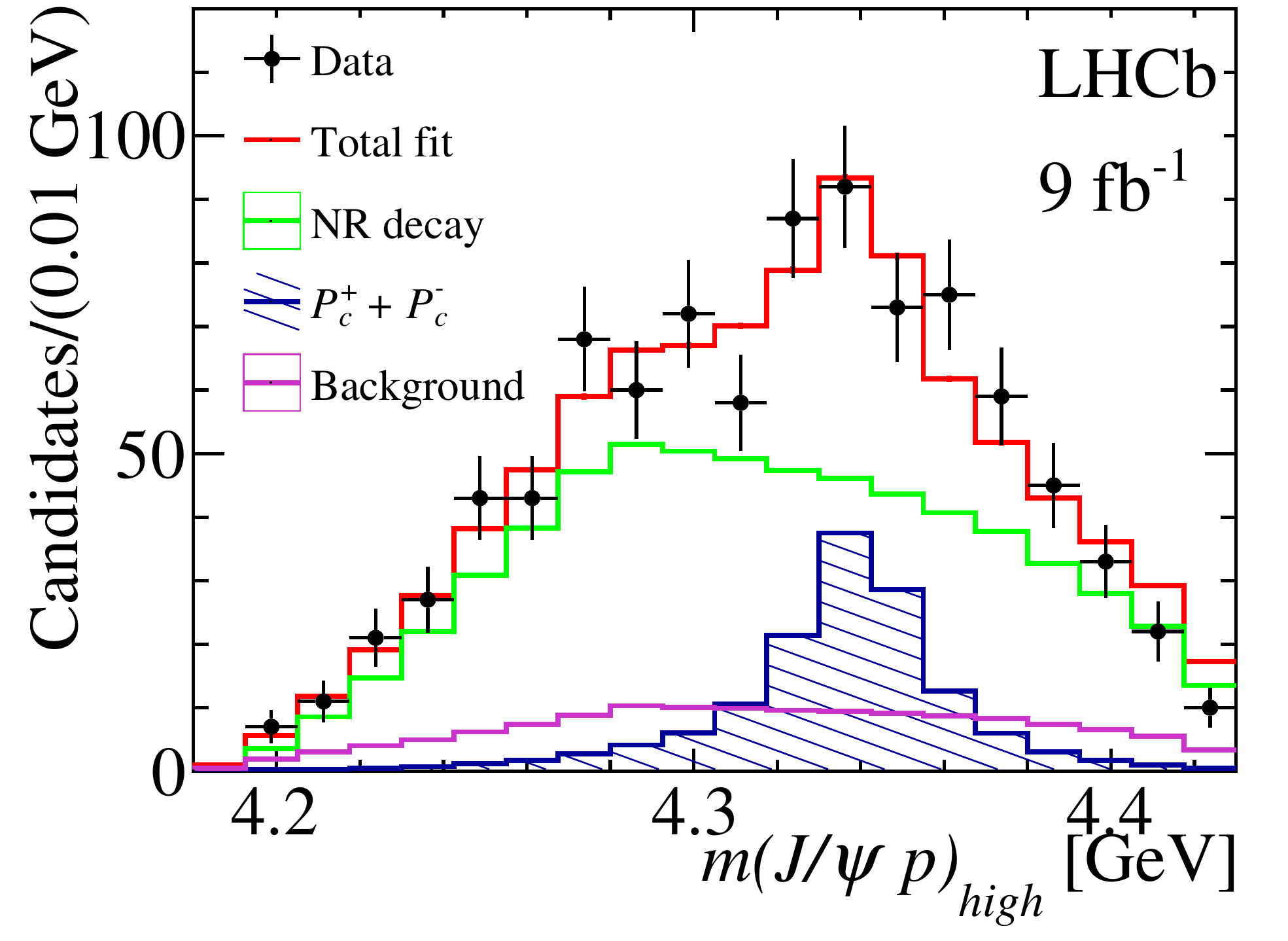}
\includegraphics[width=0.45\textwidth]{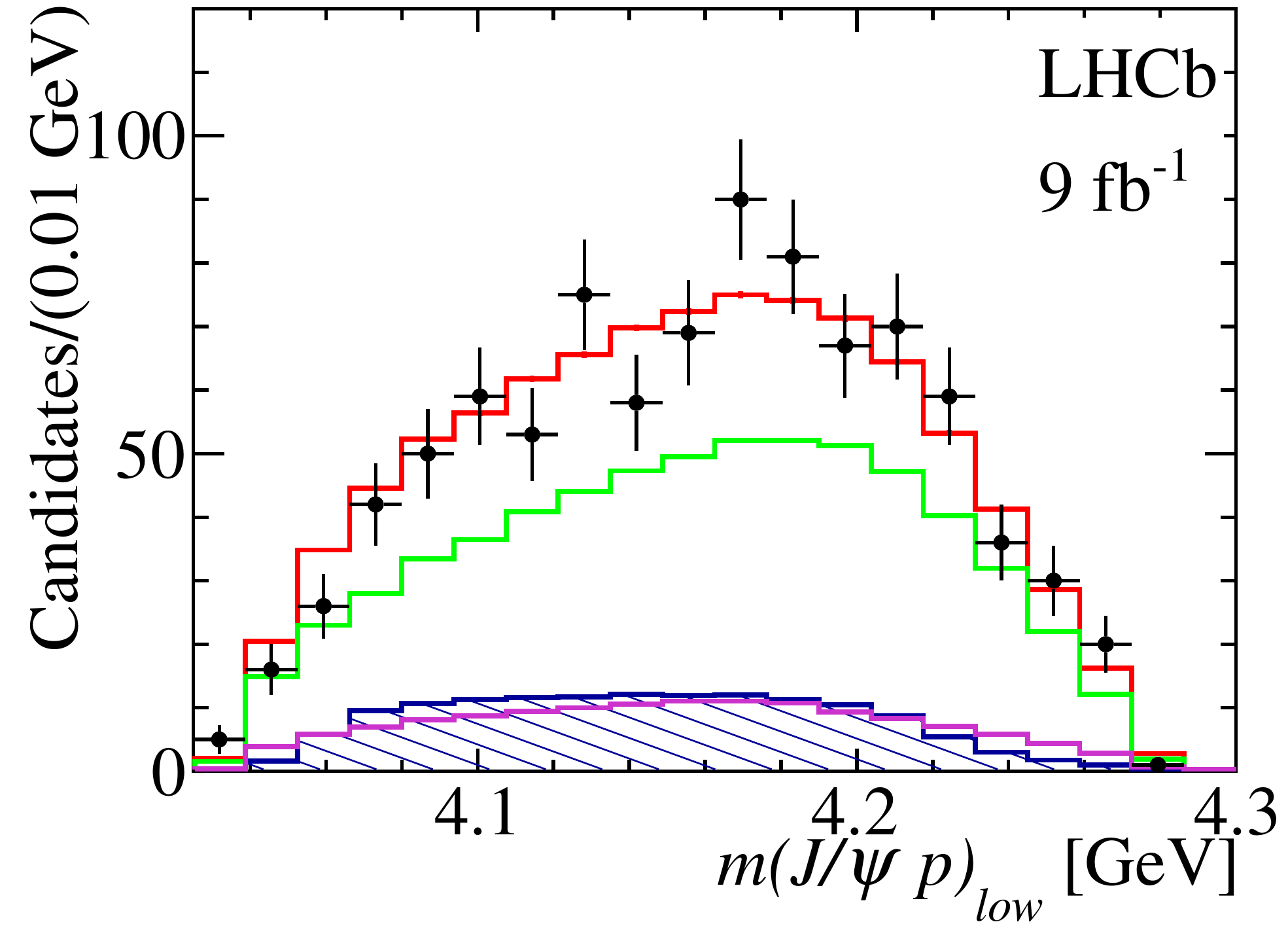}
\caption{Distribution of the maximum (left) and minimum (right) of $m(\jpsi \proton)$ and $m(\jpsi \antiproton)$. Results of the amplitude fit are superimposed, where the sum of the $P_c^+$ and $P_c^-$ contributions is shown in blue.}\label{fig:max_min_Jpsip}
\end{figure}

Since the amplitude model is symmetric for $\proton\leftrightarrow\antiproton$ inversion by construction, it is not possible to distinguish between the $P_c^+$ and $P_c^-$ states. Therefore, the distribution of the maximum and minimum of the $\jpsi\proton$ and $\jpsi \antiproton$ invariant masses, defined as
\begin{align*}
m(\jpsi\proton)_{\text{high}}&= \max(m(\jpsi\proton), m(\jpsi\antiproton)), \\ m(\jpsi\proton)_{\text{low}}&= \min(m(\jpsi\proton), m(\jpsi\antiproton)), 
\end{align*}
is shown in Fig.~\ref{fig:max_min_Jpsip}. In this way, both $P_c^+$ and $P_c^-$ contributions are visible in the distribution of $m(\jpsi \proton)_{\text{high}}$ (left), while their reflections are projected in the distribution of $m(\jpsi\proton)_{\text{low}}$ (right).

\addcontentsline{toc}{section}{References}
\setboolean{inbibliography}{true}
\bibliographystyle{LHCb}
\bibliography{main,standard,LHCb-PAPER,LHCb-CONF,LHCb-DP,LHCb-TDR}

\ifx\mcitethebibliography\mciteundefinedmacro
\PackageError{LHCb.bst}{mciteplus.sty has not been loaded}
{This bibstyle requires the use of the mciteplus package.}\fi
\providecommand{\href}[2]{#2}
\begin{mcitethebibliography}{10}
\mciteSetBstSublistMode{n}
\mciteSetBstMaxWidthForm{subitem}{\alph{mcitesubitemcount})}
\mciteSetBstSublistLabelBeginEnd{\mcitemaxwidthsubitemform\space}
{\relax}{\relax}

\bibitem{LHCb-PAPER-2015-029}
LHCb collaboration, R.~Aaij {\em et~al.},
  \ifthenelse{\boolean{articletitles}}{\emph{{Observation of $\jpsi\proton$
  resonances consistent with pentaquark states in
  \mbox{\decay{\Lb}{\jpsi\proton\Km}} decays}},
  }{}\href{https://doi.org/10.1103/PhysRevLett.115.072001}{Phys.\ Rev.\ Lett.\
  \textbf{115} (2015) 072001},
  \href{http://arxiv.org/abs/1507.03414}{{\normalfont\ttfamily
  arXiv:1507.03414}}\relax
\mciteBstWouldAddEndPuncttrue
\mciteSetBstMidEndSepPunct{\mcitedefaultmidpunct}
{\mcitedefaultendpunct}{\mcitedefaultseppunct}\relax
\EndOfBibitem
\bibitem{LHCb-PAPER-2019-014}
LHCb collaboration, R.~Aaij {\em et~al.},
  \ifthenelse{\boolean{articletitles}}{\emph{{Observation of a narrow
  pentaquark state, $P_c(4312)^+$, and of two-peak structure of the
  $P_c(4450)^+$}},
  }{}\href{https://doi.org/10.1103/PhysRevLett.122.222001}{Phys.\ Rev.\ Lett.\
  \textbf{122} (2019) 222001},
  \href{http://arxiv.org/abs/1904.03947}{{\normalfont\ttfamily
  arXiv:1904.03947}}\relax
\mciteBstWouldAddEndPuncttrue
\mciteSetBstMidEndSepPunct{\mcitedefaultmidpunct}
{\mcitedefaultendpunct}{\mcitedefaultseppunct}\relax
\EndOfBibitem
\bibitem{LHCb-PAPER-2020-039}
LHCb collaboration, R.~Aaij {\em et~al.},
  \ifthenelse{\boolean{articletitles}}{\emph{{Evidence of a $\jpsi\Lz$
  structure and observation of excited $\Xim$ states in the $\Xibm \rightarrow
  \jpsi \Lz \Km$ decay}},
  }{}\href{https://doi.org/10.1016/j.scib.2021.02.030}{Science Bulletin
  \textbf{66} (2021) 1278},
  \href{http://arxiv.org/abs/2012.10380}{{\normalfont\ttfamily
  arXiv:2012.10380}}\relax
\mciteBstWouldAddEndPuncttrue
\mciteSetBstMidEndSepPunct{\mcitedefaultmidpunct}
{\mcitedefaultendpunct}{\mcitedefaultseppunct}\relax
\EndOfBibitem
\bibitem{Esposito:2016noz}
A.~Esposito, A.~Pilloni, and A.~D. Polosa,
  \ifthenelse{\boolean{articletitles}}{\emph{{Multiquark Resonances}},
  }{}\href{https://doi.org/10.1016/j.physrep.2016.11.002}{Phys.\ Rept.\
  \textbf{668} (2017) 1},
  \href{http://arxiv.org/abs/1611.07920}{{\normalfont\ttfamily
  arXiv:1611.07920}}\relax
\mciteBstWouldAddEndPuncttrue
\mciteSetBstMidEndSepPunct{\mcitedefaultmidpunct}
{\mcitedefaultendpunct}{\mcitedefaultseppunct}\relax
\EndOfBibitem
\bibitem{Richard:2016eis}
J.-M. Richard, \ifthenelse{\boolean{articletitles}}{\emph{{Exotic hadrons:
  review and perspectives}},
  }{}\href{https://doi.org/10.1007/s00601-016-1159-0}{Few Body Syst.\
  \textbf{57} (2016) 1185},
  \href{http://arxiv.org/abs/1606.08593}{{\normalfont\ttfamily
  arXiv:1606.08593}}\relax
\mciteBstWouldAddEndPuncttrue
\mciteSetBstMidEndSepPunct{\mcitedefaultmidpunct}
{\mcitedefaultendpunct}{\mcitedefaultseppunct}\relax
\EndOfBibitem
\bibitem{Guo:2017jvc}
F.-K. Guo {\em et~al.}, \ifthenelse{\boolean{articletitles}}{\emph{{Hadronic
  molecules}}, }{}\href{https://doi.org/10.1103/RevModPhys.90.015004}{Rev.\
  Mod.\ Phys.\  \textbf{90} (2018) 015004},
  \href{http://arxiv.org/abs/1705.00141}{{\normalfont\ttfamily
  arXiv:1705.00141}}\relax
\mciteBstWouldAddEndPuncttrue
\mciteSetBstMidEndSepPunct{\mcitedefaultmidpunct}
{\mcitedefaultendpunct}{\mcitedefaultseppunct}\relax
\EndOfBibitem
\bibitem{Guo:2019twa}
F.-K. Guo, X.-H. Liu, and S.~Sakai,
  \ifthenelse{\boolean{articletitles}}{\emph{{Threshold cusps and triangle
  singularities in hadronic reactions}},
  }{}\href{https://doi.org/10.1016/j.ppnp.2020.103757}{Prog.\ Part.\ Nucl.\
  Phys.\  \textbf{112} (2020) 103757},
  \href{http://arxiv.org/abs/1912.07030}{{\normalfont\ttfamily
  arXiv:1912.07030}}\relax
\mciteBstWouldAddEndPuncttrue
\mciteSetBstMidEndSepPunct{\mcitedefaultmidpunct}
{\mcitedefaultendpunct}{\mcitedefaultseppunct}\relax
\EndOfBibitem
\bibitem{RevModPhys.90.015003}
S.~L. Olsen, T.~Skwarnicki, and D.~Zieminska,
  \ifthenelse{\boolean{articletitles}}{\emph{Nonstandard heavy mesons and
  baryons: Experimental evidence},
  }{}\href{https://doi.org/10.1103/RevModPhys.90.015003}{Rev.\ Mod.\ Phys.\
  \textbf{90} (2018) 015003}\relax
\mciteBstWouldAddEndPuncttrue
\mciteSetBstMidEndSepPunct{\mcitedefaultmidpunct}
{\mcitedefaultendpunct}{\mcitedefaultseppunct}\relax
\EndOfBibitem
\bibitem{LHCb-PAPER-2018-046}
LHCb collaboration, R.~Aaij {\em et~al.},
  \ifthenelse{\boolean{articletitles}}{\emph{{Observation of
  \mbox{\decay{\BdorBs}{\jpsi \proton\antiproton}} decays and precision
  measurements of the \BdorBs masses}},
  }{}\href{https://doi.org/10.1103/PhysRevLett.122.191804}{Phys.\ Rev.\ Lett.\
  \textbf{122} (2019) 191804},
  \href{http://arxiv.org/abs/1902.05588}{{\normalfont\ttfamily
  arXiv:1902.05588}}\relax
\mciteBstWouldAddEndPuncttrue
\mciteSetBstMidEndSepPunct{\mcitedefaultmidpunct}
{\mcitedefaultendpunct}{\mcitedefaultseppunct}\relax
\EndOfBibitem
\bibitem{Hsiao:2014tda}
Y.~K. Hsiao and C.~Q. Geng,
  \ifthenelse{\boolean{articletitles}}{\emph{{$f_J(2220)$ and hadronic
  $\bar{B}^0_s$ decays}},
  }{}\href{https://doi.org/10.1140/epjc/s10052-015-3317-9}{Eur.\ Phys.\ J.\
  \textbf{C75} (2015) 101},
  \href{http://arxiv.org/abs/1412.4900}{{\normalfont\ttfamily
  arXiv:1412.4900}}\relax
\mciteBstWouldAddEndPuncttrue
\mciteSetBstMidEndSepPunct{\mcitedefaultmidpunct}
{\mcitedefaultendpunct}{\mcitedefaultseppunct}\relax
\EndOfBibitem
\bibitem{Rosner_2003}
J.~L. Rosner, \ifthenelse{\boolean{articletitles}}{\emph{{Low mass baryon
  anti-baryon enhancements in B decays}},
  }{}\href{https://doi.org/10.1103/PhysRevD.68.014004}{Phys.\ Rev.\
  \textbf{D68} (2003) 014004},
  \href{http://arxiv.org/abs/hep-ph/0303079}{{\normalfont\ttfamily
  arXiv:hep-ph/0303079}}\relax
\mciteBstWouldAddEndPuncttrue
\mciteSetBstMidEndSepPunct{\mcitedefaultmidpunct}
{\mcitedefaultendpunct}{\mcitedefaultseppunct}\relax
\EndOfBibitem
\bibitem{LHCb-DP-2014-002}
LHCb collaboration, R.~Aaij {\em et~al.},
  \ifthenelse{\boolean{articletitles}}{\emph{{LHCb detector performance}},
  }{}\href{https://doi.org/10.1142/S0217751X15300227}{Int.\ J.\ Mod.\ Phys.\
  \textbf{A30} (2015) 1530022},
  \href{http://arxiv.org/abs/1412.6352}{{\normalfont\ttfamily
  arXiv:1412.6352}}\relax
\mciteBstWouldAddEndPuncttrue
\mciteSetBstMidEndSepPunct{\mcitedefaultmidpunct}
{\mcitedefaultendpunct}{\mcitedefaultseppunct}\relax
\EndOfBibitem
\bibitem{LHCb-DP-2014-001}
R.~Aaij {\em et~al.}, \ifthenelse{\boolean{articletitles}}{\emph{{Performance
  of the LHCb Vertex Locator}},
  }{}\href{https://doi.org/10.1088/1748-0221/9/09/P09007}{JINST \textbf{9}
  (2014) P09007}, \href{http://arxiv.org/abs/1405.7808}{{\normalfont\ttfamily
  arXiv:1405.7808}}\relax
\mciteBstWouldAddEndPuncttrue
\mciteSetBstMidEndSepPunct{\mcitedefaultmidpunct}
{\mcitedefaultendpunct}{\mcitedefaultseppunct}\relax
\EndOfBibitem
\bibitem{LHCb-DP-2013-003}
R.~Arink {\em et~al.}, \ifthenelse{\boolean{articletitles}}{\emph{{Performance
  of the LHCb Outer Tracker}},
  }{}\href{https://doi.org/10.1088/1748-0221/9/01/P01002}{JINST \textbf{9}
  (2014) P01002}, \href{http://arxiv.org/abs/1311.3893}{{\normalfont\ttfamily
  arXiv:1311.3893}}\relax
\mciteBstWouldAddEndPuncttrue
\mciteSetBstMidEndSepPunct{\mcitedefaultmidpunct}
{\mcitedefaultendpunct}{\mcitedefaultseppunct}\relax
\EndOfBibitem
\bibitem{LHCb-DP-2012-002}
A.~A. Alves~Jr.\ {\em et~al.},
  \ifthenelse{\boolean{articletitles}}{\emph{{Performance of the LHCb muon
  system}}, }{}\href{https://doi.org/10.1088/1748-0221/8/02/P02022}{JINST
  \textbf{8} (2013) P02022},
  \href{http://arxiv.org/abs/1211.1346}{{\normalfont\ttfamily
  arXiv:1211.1346}}\relax
\mciteBstWouldAddEndPuncttrue
\mciteSetBstMidEndSepPunct{\mcitedefaultmidpunct}
{\mcitedefaultendpunct}{\mcitedefaultseppunct}\relax
\EndOfBibitem
\bibitem{LHCb-DP-2012-004}
R.~Aaij {\em et~al.}, \ifthenelse{\boolean{articletitles}}{\emph{{The \lhcb
  trigger and its performance in 2011}},
  }{}\href{https://doi.org/10.1088/1748-0221/8/04/P04022}{JINST \textbf{8}
  (2013) P04022}, \href{http://arxiv.org/abs/1211.3055}{{\normalfont\ttfamily
  arXiv:1211.3055}}\relax
\mciteBstWouldAddEndPuncttrue
\mciteSetBstMidEndSepPunct{\mcitedefaultmidpunct}
{\mcitedefaultendpunct}{\mcitedefaultseppunct}\relax
\EndOfBibitem
\bibitem{Sjostrand:2007gs}
T.~Sj\"{o}strand, S.~Mrenna, and P.~Skands,
  \ifthenelse{\boolean{articletitles}}{\emph{{A brief introduction to PYTHIA
  8.1}}, }{}\href{https://doi.org/10.1016/j.cpc.2008.01.036}{Comput.\ Phys.\
  Commun.\  \textbf{178} (2008) 852},
  \href{http://arxiv.org/abs/0710.3820}{{\normalfont\ttfamily
  arXiv:0710.3820}}\relax
\mciteBstWouldAddEndPuncttrue
\mciteSetBstMidEndSepPunct{\mcitedefaultmidpunct}
{\mcitedefaultendpunct}{\mcitedefaultseppunct}\relax
\EndOfBibitem
\bibitem{LHCb-PROC-2010-056}
I.~Belyaev {\em et~al.}, \ifthenelse{\boolean{articletitles}}{\emph{{Handling
  of the generation of primary events in Gauss, the LHCb simulation
  framework}}, }{}\href{https://doi.org/10.1088/1742-6596/331/3/032047}{J.\
  Phys.\ Conf.\ Ser.\  \textbf{331} (2011) 032047}\relax
\mciteBstWouldAddEndPuncttrue
\mciteSetBstMidEndSepPunct{\mcitedefaultmidpunct}
{\mcitedefaultendpunct}{\mcitedefaultseppunct}\relax
\EndOfBibitem
\bibitem{Lange:2001uf}
D.~J. Lange, \ifthenelse{\boolean{articletitles}}{\emph{{The EvtGen particle
  decay simulation package}},
  }{}\href{https://doi.org/10.1016/S0168-9002(01)00089-4}{Nucl.\ Instrum.\
  Meth.\  \textbf{A462} (2001) 152}\relax
\mciteBstWouldAddEndPuncttrue
\mciteSetBstMidEndSepPunct{\mcitedefaultmidpunct}
{\mcitedefaultendpunct}{\mcitedefaultseppunct}\relax
\EndOfBibitem
\bibitem{davidson2015photos}
N.~Davidson, T.~Przedzinski, and Z.~Was,
  \ifthenelse{\boolean{articletitles}}{\emph{{PHOTOS interface in C++:
  Technical and physics documentation}},
  }{}\href{https://doi.org/https://doi.org/10.1016/j.cpc.2015.09.013}{Comp.\
  Phys.\ Comm.\  \textbf{199} (2016) 86},
  \href{http://arxiv.org/abs/1011.0937}{{\normalfont\ttfamily
  arXiv:1011.0937}}\relax
\mciteBstWouldAddEndPuncttrue
\mciteSetBstMidEndSepPunct{\mcitedefaultmidpunct}
{\mcitedefaultendpunct}{\mcitedefaultseppunct}\relax
\EndOfBibitem
\bibitem{Allison:2006ve}
Geant4 collaboration, J.~Allison {\em et~al.},
  \ifthenelse{\boolean{articletitles}}{\emph{{Geant4 developments and
  applications}}, }{}\href{https://doi.org/10.1109/TNS.2006.869826}{IEEE
  Trans.\ Nucl.\ Sci.\  \textbf{53} (2006) 270}\relax
\mciteBstWouldAddEndPuncttrue
\mciteSetBstMidEndSepPunct{\mcitedefaultmidpunct}
{\mcitedefaultendpunct}{\mcitedefaultseppunct}\relax
\EndOfBibitem
\bibitem{Agostinelli:2002hh}
Geant4 collaboration, S.~Agostinelli {\em et~al.},
  \ifthenelse{\boolean{articletitles}}{\emph{{Geant4: A simulation toolkit}},
  }{}\href{https://doi.org/10.1016/S0168-9002(03)01368-8}{Nucl.\ Instrum.\
  Meth.\  \textbf{A506} (2003) 250}\relax
\mciteBstWouldAddEndPuncttrue
\mciteSetBstMidEndSepPunct{\mcitedefaultmidpunct}
{\mcitedefaultendpunct}{\mcitedefaultseppunct}\relax
\EndOfBibitem
\bibitem{LHCb-PROC-2011-006}
M.~Clemencic {\em et~al.}, \ifthenelse{\boolean{articletitles}}{\emph{{The
  \lhcb simulation application, Gauss: Design, evolution and experience}},
  }{}\href{https://doi.org/10.1088/1742-6596/331/3/032023}{J.\ Phys.\ Conf.\
  Ser.\  \textbf{331} (2011) 032023}\relax
\mciteBstWouldAddEndPuncttrue
\mciteSetBstMidEndSepPunct{\mcitedefaultmidpunct}
{\mcitedefaultendpunct}{\mcitedefaultseppunct}\relax
\EndOfBibitem
\bibitem{LHCb-PAPER-2014-059}
LHCb collaboration, R.~Aaij {\em et~al.},
  \ifthenelse{\boolean{articletitles}}{\emph{{Precision measurement of \CP
  violation in \mbox{\decay{\Bs}{\jpsi\Kp\Km}} decays}},
  }{}\href{https://doi.org/10.1103/PhysRevLett.114.041801}{Phys.\ Rev.\ Lett.\
  \textbf{114} (2015) 041801},
  \href{http://arxiv.org/abs/1411.3104}{{\normalfont\ttfamily
  arXiv:1411.3104}}\relax
\mciteBstWouldAddEndPuncttrue
\mciteSetBstMidEndSepPunct{\mcitedefaultmidpunct}
{\mcitedefaultendpunct}{\mcitedefaultseppunct}\relax
\EndOfBibitem
\bibitem{Hulsbergen:2005pu}
W.~D. Hulsbergen, \ifthenelse{\boolean{articletitles}}{\emph{{Decay chain
  fitting with a Kalman filter}},
  }{}\href{https://doi.org/10.1016/j.nima.2005.06.078}{Nucl.\ Instrum.\ Meth.\
  \textbf{A552} (2005) 566},
  \href{http://arxiv.org/abs/physics/0503191}{{\normalfont\ttfamily
  arXiv:physics/0503191}}\relax
\mciteBstWouldAddEndPuncttrue
\mciteSetBstMidEndSepPunct{\mcitedefaultmidpunct}
{\mcitedefaultendpunct}{\mcitedefaultseppunct}\relax
\EndOfBibitem
\bibitem{PDG20}
Particle Data Group, P.~A. Zyla {\em et~al.},
  \ifthenelse{\boolean{articletitles}}{\emph{{Review of Particle Physics}},
  }{}\href{https://doi.org/10.1093/ptep/ptaa104}{Prog.\ Theor.\ Exp.\ Phys.\
  \textbf{2020} (2020) 083C01}\relax
\mciteBstWouldAddEndPuncttrue
\mciteSetBstMidEndSepPunct{\mcitedefaultmidpunct}
{\mcitedefaultendpunct}{\mcitedefaultseppunct}\relax
\EndOfBibitem
\bibitem{Breiman}
L.~Breiman, J.~H. Friedman, R.~A. Olshen, and C.~J. Stone, {\em Classification
  and regression trees}, Wadsworth international group, Belmont, California,
  USA, 1984\relax
\mciteBstWouldAddEndPuncttrue
\mciteSetBstMidEndSepPunct{\mcitedefaultmidpunct}
{\mcitedefaultendpunct}{\mcitedefaultseppunct}\relax
\EndOfBibitem
\bibitem{LHCb-DP-2018-001}
R.~Aaij {\em et~al.}, \ifthenelse{\boolean{articletitles}}{\emph{{Selection and
  processing of calibration samples to measure the particle identification
  performance of the LHCb experiment in Run 2}},
  }{}\href{https://doi.org/10.1140/epjti/s40485-019-0050-z}{Eur.\ Phys.\ J.\
  Tech.\ Instr.\  \textbf{6} (2018) 1},
  \href{http://arxiv.org/abs/1803.00824}{{\normalfont\ttfamily
  arXiv:1803.00824}}\relax
\mciteBstWouldAddEndPuncttrue
\mciteSetBstMidEndSepPunct{\mcitedefaultmidpunct}
{\mcitedefaultendpunct}{\mcitedefaultseppunct}\relax
\EndOfBibitem
\bibitem{Skwarnicki:1986xj}
T.~Skwarnicki, {\em {A study of the radiative cascade transitions between the
  Upsilon-prime and Upsilon resonances}}, PhD thesis, Institute of Nuclear
  Physics, Krakow, 1986,
  {\href{http://inspirehep.net/record/230779/}{DESY-F31-86-02}}\relax
\mciteBstWouldAddEndPuncttrue
\mciteSetBstMidEndSepPunct{\mcitedefaultmidpunct}
{\mcitedefaultendpunct}{\mcitedefaultseppunct}\relax
\EndOfBibitem
\bibitem{Chung:186421}
S.~U. Chung, \ifthenelse{\boolean{articletitles}}{\emph{{Spin formalisms}}, }{}
  1971.
\newblock CERN, Geneva, 1969 - 1970,
  doi:~\href{https://doi.org/10.5170/CERN-1971-008}{10.5170/CERN-1971-008}\relax
\mciteBstWouldAddEndPuncttrue
\mciteSetBstMidEndSepPunct{\mcitedefaultmidpunct}
{\mcitedefaultendpunct}{\mcitedefaultseppunct}\relax
\EndOfBibitem
\bibitem{Jackob1959:at}
M.~Jacob and G.~C. Wick, \ifthenelse{\boolean{articletitles}}{\emph{{On the
  General Theory of Collisions for Particles with Spin}},
  }{}\href{https://doi.org/10.1016/0003-4916(59)90051-X}{Annals of Physics
  \textbf{7} (1959) 404}\relax
\mciteBstWouldAddEndPuncttrue
\mciteSetBstMidEndSepPunct{\mcitedefaultmidpunct}
{\mcitedefaultendpunct}{\mcitedefaultseppunct}\relax
\EndOfBibitem
\bibitem{DPdeco}
JPAC collaboration, M.~Mikhasenko {\em et~al.},
  \ifthenelse{\boolean{articletitles}}{\emph{{Dalitz-plot decomposition for
  three-body decays}},
  }{}\href{https://doi.org/10.1103/PhysRevD.101.034033}{Phys.\ Rev.\
  \textbf{D101} (2020) 034033},
  \href{http://arxiv.org/abs/1910.04566}{{\normalfont\ttfamily
  arXiv:1910.04566}}\relax
\mciteBstWouldAddEndPuncttrue
\mciteSetBstMidEndSepPunct{\mcitedefaultmidpunct}
{\mcitedefaultendpunct}{\mcitedefaultseppunct}\relax
\EndOfBibitem
\bibitem{PhysRevLett.76.3502}
BES collaboration, J.~Z. Bai {\em et~al.},
  \ifthenelse{\boolean{articletitles}}{\emph{Studies of
  $\ensuremath{\xi}$(2230) in $\mathit{{J}}$/$\ensuremath{\psi}$ radiative
  decays}, }{}\href{https://doi.org/10.1103/PhysRevLett.76.3502}{Phys.\ Rev.\
  Lett.\  \textbf{76} (1996) 3502}\relax
\mciteBstWouldAddEndPuncttrue
\mciteSetBstMidEndSepPunct{\mcitedefaultmidpunct}
{\mcitedefaultendpunct}{\mcitedefaultseppunct}\relax
\EndOfBibitem
\bibitem{Ablikim:2005um}
BES collaboration, M.~Ablikim {\em et~al.},
  \ifthenelse{\boolean{articletitles}}{\emph{{Observation of a resonance
  X(1835) in $J/\psi \to \gamma \pi^+ \pi^- \eta^{\prime}$}},
  }{}\href{https://doi.org/10.1103/PhysRevLett.95.262001}{Phys.\ Rev.\ Lett.\
  \textbf{95} (2005) 262001},
  \href{http://arxiv.org/abs/hep-ex/0508025}{{\normalfont\ttfamily
  arXiv:hep-ex/0508025}}\relax
\mciteBstWouldAddEndPuncttrue
\mciteSetBstMidEndSepPunct{\mcitedefaultmidpunct}
{\mcitedefaultendpunct}{\mcitedefaultseppunct}\relax
\EndOfBibitem
\bibitem{BESIII:2011aa}
BESIII collaboration, M.~Ablikim {\em et~al.},
  \ifthenelse{\boolean{articletitles}}{\emph{{Spin-Parity Analysis of
  $p\bar{p}$ Mass Threshold Structure in $J/\psi$ and $\psi^\prime$ Radiative
  Decays}}, }{}\href{https://doi.org/10.1103/PhysRevLett.108.112003}{Phys.\
  Rev.\ Lett.\  \textbf{108} (2012) 112003},
  \href{http://arxiv.org/abs/1112.0942}{{\normalfont\ttfamily
  arXiv:1112.0942}}\relax
\mciteBstWouldAddEndPuncttrue
\mciteSetBstMidEndSepPunct{\mcitedefaultmidpunct}
{\mcitedefaultendpunct}{\mcitedefaultseppunct}\relax
\EndOfBibitem
\bibitem{LHCb-PAPER-2015-014}
LHCb collaboration, R.~Aaij {\em et~al.},
  \ifthenelse{\boolean{articletitles}}{\emph{{A study of \CP violation in
  \mbox{\decay{\Bmp}{\D h^\mp}} $(h=K,\pi)$ with the modes
  \mbox{\decay{\D}{\Kmp\pipm\piz}}, \mbox{\decay{\D}{\pip\pim\piz}} and
  \mbox{\decay{\D}{\Kp\Km\piz}}}},
  }{}\href{https://doi.org/10.1103/PhysRevD.91.112014}{Phys.\ Rev.\
  \textbf{D91} (2015) 112014},
  \href{http://arxiv.org/abs/1504.05442}{{\normalfont\ttfamily
  arXiv:1504.05442}}\relax
\mciteBstWouldAddEndPuncttrue
\mciteSetBstMidEndSepPunct{\mcitedefaultmidpunct}
{\mcitedefaultendpunct}{\mcitedefaultseppunct}\relax
\EndOfBibitem
\bibitem{marangotto2020helicity}
D.~Marangotto, \ifthenelse{\boolean{articletitles}}{\emph{{Helicity Amplitudes
  for Generic Multibody Particle Decays Featuring Multiple Decay Chains}},
  }{}\href{https://doi.org/10.1155/2020/6674595}{Adv.\ High Energy Phys.\
  \textbf{2020} (2020) 6674595},
  \href{http://arxiv.org/abs/1911.10025}{{\normalfont\ttfamily
  arXiv:1911.10025}}\relax
\mciteBstWouldAddEndPuncttrue
\mciteSetBstMidEndSepPunct{\mcitedefaultmidpunct}
{\mcitedefaultendpunct}{\mcitedefaultseppunct}\relax
\EndOfBibitem
\bibitem{wang2020novel}
M.~Wang {\em et~al.}, \ifthenelse{\boolean{articletitles}}{\emph{{A novel
  method to test particle ordering and final state alignment in helicity
  formalism}}, }{}\href{https://doi.org/10.1088/1674-1137/abf139}{Chin.\ Phys.\
   \textbf{C45} (2021) 063103},
  \href{http://arxiv.org/abs/2012.03699}{{\normalfont\ttfamily
  arXiv:2012.03699}}\relax
\mciteBstWouldAddEndPuncttrue
\mciteSetBstMidEndSepPunct{\mcitedefaultmidpunct}
{\mcitedefaultendpunct}{\mcitedefaultseppunct}\relax
\EndOfBibitem
\bibitem{CLs}
A.~L. Read, \ifthenelse{\boolean{articletitles}}{\emph{{Presentation of search
  results: The CL$_{\rm s}$ technique}},
  }{}\href{https://doi.org/10.1088/0954-3899/28/10/313}{J.\ Phys.\
  \textbf{G28} (2002) 2693}\relax
\mciteBstWouldAddEndPuncttrue
\mciteSetBstMidEndSepPunct{\mcitedefaultmidpunct}
{\mcitedefaultendpunct}{\mcitedefaultseppunct}\relax
\EndOfBibitem
\bibitem{doi:10.1142/6096}
F.~James, {\em Statistical Methods in Experimental Physics},
  \href{https://doi.org/10.1142/6096}{ World Scientific, 2nd~ed., 2006}\relax
\mciteBstWouldAddEndPuncttrue
\mciteSetBstMidEndSepPunct{\mcitedefaultmidpunct}
{\mcitedefaultendpunct}{\mcitedefaultseppunct}\relax
\EndOfBibitem
\bibitem{Wu_2014}
N.~Wu,
  \ifthenelse{\boolean{articletitles}}{\emph{Centrifugal{\textemdash}barrier
  effects and determination of interaction radius},
  }{}\href{https://doi.org/10.1088/0253-6102/61/1/14}{Communications in
  Theoretical Physics \textbf{61} (2014) 89}\relax
\mciteBstWouldAddEndPuncttrue
\mciteSetBstMidEndSepPunct{\mcitedefaultmidpunct}
{\mcitedefaultendpunct}{\mcitedefaultseppunct}\relax
\EndOfBibitem
\bibitem{Shen:2017ayv}
C.-W. Shen, D.~R\"onchen, U.-G. Mei\ss{}ner, and B.-S. Zou,
  \ifthenelse{\boolean{articletitles}}{\emph{{Exploratory study of possible
  resonances in heavy meson - heavy baryon coupled-channel interactions}},
  }{}\href{https://doi.org/10.1088/1674-1137/42/2/023106}{Chin.\ Phys.\ C
  \textbf{42} (2018) 023106},
  \href{http://arxiv.org/abs/1710.03885}{{\normalfont\ttfamily
  arXiv:1710.03885}}\relax
\mciteBstWouldAddEndPuncttrue
\mciteSetBstMidEndSepPunct{\mcitedefaultmidpunct}
{\mcitedefaultendpunct}{\mcitedefaultseppunct}\relax
\EndOfBibitem
\bibitem{Abe:2002tw}
Belle collaboration, K.~Abe {\em et~al.},
  \ifthenelse{\boolean{articletitles}}{\emph{{Observation of $\Bdb \to D^{(*)0}
  \proton \antiproton$}},
  }{}\href{https://doi.org/10.1103/PhysRevLett.89.151802}{Phys.\ Rev.\ Lett.\
  \textbf{89} (2002) 151802},
  \href{http://arxiv.org/abs/hep-ex/0205083}{{\normalfont\ttfamily
  arXiv:hep-ex/0205083}}\relax
\mciteBstWouldAddEndPuncttrue
\mciteSetBstMidEndSepPunct{\mcitedefaultmidpunct}
{\mcitedefaultendpunct}{\mcitedefaultseppunct}\relax
\EndOfBibitem
\bibitem{PhysRevLett.92.131801}
Belle collaboration, M.~Z. Wang {\em et~al.},
  \ifthenelse{\boolean{articletitles}}{\emph{Observation of
  ${B}^{+}\ensuremath{\rightarrow}p\overline{p}{\ensuremath{\pi}}^{+}$,
  ${B}^{0}\ensuremath{\rightarrow}p\overline{p}{K}^{0}$, and
  ${B}^{+}\ensuremath{\rightarrow}p\overline{p}{K}^{*+}$},
  }{}\href{https://doi.org/10.1103/PhysRevLett.92.131801}{Phys.\ Rev.\ Lett.\
  \textbf{92} (2004) 131801},
  \href{http://arxiv.org/abs/hep-ex/0310018}{{\normalfont\ttfamily
  arXiv:hep-ex/0310018}}\relax
\mciteBstWouldAddEndPuncttrue
\mciteSetBstMidEndSepPunct{\mcitedefaultmidpunct}
{\mcitedefaultendpunct}{\mcitedefaultseppunct}\relax
\EndOfBibitem
\bibitem{PhysRevD.72.051101}
BaBar collaboration, B.~Aubert {\em et~al.},
  \ifthenelse{\boolean{articletitles}}{\emph{Measurement of the
  ${B}^{+}\ensuremath{\rightarrow}p\overline{p}{K}^{+}$ branching fraction and
  study of the decay dynamics},
  }{}\href{https://doi.org/10.1103/PhysRevD.72.051101}{Phys.\ Rev.\
  \textbf{D72} (2005) 051101},
  \href{http://arxiv.org/abs/hep-ex/0507012}{{\normalfont\ttfamily
  arXiv:hep-ex/0507012}}\relax
\mciteBstWouldAddEndPuncttrue
\mciteSetBstMidEndSepPunct{\mcitedefaultmidpunct}
{\mcitedefaultendpunct}{\mcitedefaultseppunct}\relax
\EndOfBibitem
\bibitem{LHCb-PAPER-2012-047}
LHCb collaboration, R.~Aaij {\em et~al.},
  \ifthenelse{\boolean{articletitles}}{\emph{{Measurements of the branching
  fractions of \mbox{\decay{\Bp}{\proton\antiproton\Kp}} decays}},
  }{}\href{https://doi.org/10.1140/epjc/s10052-013-2462-2}{Eur.\ Phys.\ J.\
  \textbf{C73} (2013) 2462},
  \href{http://arxiv.org/abs/1303.7133}{{\normalfont\ttfamily
  arXiv:1303.7133}}\relax
\mciteBstWouldAddEndPuncttrue
\mciteSetBstMidEndSepPunct{\mcitedefaultmidpunct}
{\mcitedefaultendpunct}{\mcitedefaultseppunct}\relax
\EndOfBibitem
\end{mcitethebibliography}

\newpage
\centerline
{\large\bf LHCb collaboration}
\begin
{flushleft}
\small
R.~Aaij$^{32}$,
A.S.W.~Abdelmotteleb$^{56}$,
C.~Abell{\'a}n~Beteta$^{50}$,
T.~Ackernley$^{60}$,
B.~Adeva$^{46}$,
M.~Adinolfi$^{54}$,
H.~Afsharnia$^{9}$,
C.A.~Aidala$^{86}$,
S.~Aiola$^{25}$,
Z.~Ajaltouni$^{9}$
% ...

{\footnotesize \it

% ...
$^{9}$Universit{\'e} Clermont Auvergne, CNRS/IN2P3, LPC, Clermont-Ferrand, France\\
% ...
$^{25}$INFN Sezione di Milano, Milano, Italy\\
% ...
$^{32}$Nikhef National Institute for Subatomic Physics, Amsterdam, Netherlands\\
% ...
$^{46}$Instituto Galego de F{\'\i}sica de Altas Enerx{\'\i}as (IGFAE), Universidade de Santiago de Compostela, Santiago de Compostela, Spain\\
% ...
$^{50}$Physik-Institut, Universit{\"a}t Z{\"u}rich, Z{\"u}rich, Switzerland\\
% ...
$^{54}$H.H. Wills Physics Laboratory, University of Bristol, Bristol, United Kingdom\\
% ...
$^{56}$Department of Physics, University of Warwick, Coventry, United Kingdom\\
% ...
$^{60}$Oliver Lodge Laboratory, University of Liverpool, Liverpool, United Kingdom\\
% ...
$^{68}$Syracuse University, Syracuse, NY, United States\\
% ...
$^{86}$University of Michigan, Ann Arbor, United States, associated to $^{68}$\\
\bigskip
% ...
\medskip
}
\end{flushleft}

\end{document}